\documentclass[usenatbib,useAMS,manuscript]{mn2e}
\usepackage{amssymb}
\usepackage{graphicx}
\usepackage{multirow}

\newcommand{\beq}{\begin{equation}}              
\newcommand{\eeq}{\end{equation}}                
\newcommand{\Ri}{\langle R_{i} \rangle}
\newcommand{\twoRi}{\langle 2 R_{i} \rangle}
\newcommand{\LCDM}{\Lambda{\rm CDM}}
\newcommand{\kms}{{{\rm \,km~s^{-1}}}}
\newcommand{\Mvir}{M_{\rm vir}}
\newcommand{\Mstars}{M_{\rm stars}}
\newcommand{\Msun}{{\rm M}_{\odot}}
\newcommand{\vc}{v_{\rm c}}
\newcommand{\vcstar}{v_{\rm c}^*}

\title
[Galaxy evolution from strong lensing]
{Galaxy evolution from strong lensing statistics: the differential evolution of
the velocity dispersion function in concord with the $\LCDM$ paradigm}
\author[K.-H. Chae]{
Kyu-Hyun Chae\thanks{chae@sejong.ac.kr} \\
{Sejong University, Department of Astronomy and Space Science, 98
Gunja-dong, Gwangjin-Gu, Seoul 143-747, Republic of Korea}
}
\date{
Accepted ........
Received .......;
in original form ......}

\pubyear{2010}

\begin{document}
\maketitle

\begin{abstract}

We study galaxy evolution from $z=1$ to $z=0$ as a function of velocity
dispersion $\sigma$ for galaxies with $\sigma \ga 95 \kms$ based on the 
measured and Monte Carlo realised local velocity dispersion functions (VDFs) 
of galaxies and the revised statistical properties 
of 30 strongly-lensed sources from the Cosmic Lens All-Sky Survey (CLASS), the 
PMN-NVSS Extragalactic Lens Survey (PANELS) and the Hubble Space Telescope
Snapshot survey. We assume that the total (luminous plus dark) mass profile
of a galaxy is isothermal in the optical region for $0 \le z \le 1$ as 
suggested by mass modelling of lensing galaxies.
 This study is the first to investigate the evolution of the VDF shape 
as well as the overall number density. It is also the first to
study the evolution of the total and the late-type VDFs in addition to the 
early-type VDF. For the evolutionary behaviours of the VDFs we find that: 
(1) the number density of massive (mostly early-type) galaxies with 
$\sigma \ga 200 \kms$ evolves differentially in the way that the number density
evolution is greater at a higher velocity dispersion; 
(2) the number density of intermediate and low mass early-type galaxies 
($95 \kms \la \sigma \la 200 \kms$) is nearly constant; 
(3) the late-type VDF transformed from the Monte Carlo realised 
circular velocity function is consistent with no evolution in its shape or
integrated number density consistent with galaxy survey results. 
These evolutionary behaviours of the VDFs are strikingly similar to those of 
the dark halo mass function (DMF) from N-body simulations and the stellar mass 
function (SMF) predicted by recent semi-analytic models of galaxy formation 
under the current $\LCDM$ hierarchical structure formation paradigm. 
Interestingly, the VDF evolutions appear to be qualitatively different from 
``stellar mass-downsizing'' evolutions obtained by many galaxy surveys. 
The coevolution of the DMF, the VDF and the SMF is investigated in quantitative
detail based on up-to-date theoretical and observational results in a following
paper. We consider several possible  systematic errors for the lensing analysis
 and find that they are not likely to alter the conclusions. 
\end{abstract}

\maketitle

\section{Introduction}

The observationally-derived statistical 
properties of galaxies provide key constraints on models of galaxy formation 
and evolution. The statistical properties of galaxies include the luminosity
functions (LFs), the stellar mass functions (SMFs) and the velocity 
functions (VFs). In the current Lambda cold dark matter ($\LCDM$) 
hierarchical structure formation picture CDM haloes are formed due to 
gravitational instabilities and evolve hierarchically through merging
(e.g.\ \citealt{WR78,LC93}).
Baryons settle in the dark matter halo potential wells and undergo 
dissipational radiative processes including star formations and
evolutions, supernovae explosions, AGN activities and their feedback,
 resulting in visible galaxies (in the central parts of
the haloes). Visible galaxies may further merge (and become 
morphologically transformed) resulting in evolutions of galaxy populations.
The various statistical functions of galaxies are the end results of complex
processes involving gravitational physics, baryonic physics and merging.

The LFs of galaxies have traditionally been measured most extensively
and reliably (e.g.\ \citealt{Col01,Koc01,Nor02,Bla03,Cro05}) and 
consequently have provided vital constraints on models of galaxy 
formation and evolution. More recently, the SMFs have been 
the focus of many observational studies (e.g.\ 
\citealt{Col01,Bel03,Dro04,Bor06,Bun06,Cim06,Fon06,Poz07,Con07,Sca07,Mar09}) 
because they can provide more robust tests of the underlying $\LCDM$ theory. 
Recent semi-analytic models of galaxy formation and evolution 
(e.g.\ \citealt{Cro06,Bow06,KW07,Str08}) appear to match the observed 
local LFs/SMFs owing in large part to the suitable incorporation of baryonic 
physics such as AGN feedback.  

However, the evolutions of the LFs/SMFs from a high redshift Universe are not
 understood as well. On the observational side, deep surveys of galaxies 
have been used to derive the evolutions for the type-specific
populations as well as the total population, particular attention paid to
photometrically red galaxies or morphologically early-type galaxies which
occupy the most massive part of the galaxy population. Various existing
observational results on the number density evolution are at variance. 
Many galaxy survey results
(e.g. \citealt{Cim06,Bun06,Fon06,Poz07,Con07,Sca07,Bro07,Coo08,Mar09})
 suggest little evolution of massive (super-$L^*$) early-type galaxies often 
accompanying ``mass-downsizing'' behaviour  in which more massive galaxies 
are to be in place at an earlier time so that the number density of most 
massive early-type galaxies evolves least while the number density of
 typical $\sim L^*$ galaxies may evolve significantly over cosmic time. 
On the other hand, there are also galaxy survey results 
that do not particularly support mass-downsizing evolutions 
(e.g.\ \citealt{Bel04,Ilb06,Fab07,Lot08,Ilb09}).
On the theoretical side, semi-analytic model predictions on the LF/SMF 
evolutions scatter (e.g.\ 
\citealt{Bow06,Men06,Mon06,deL06,KW07,Alm08,Cat08,Str08,Fon09}) 
and are not in good agreement with existing observational constraints on 
the evolutions of the SMFs, particularly for the type-specific galaxy 
populations. Semi-analytic models do not predict  mass-downsizing evolutions
of the SMF and according to more recent studies the mismatch between model 
predictions and observed SMFs is more severe for less massive galaxies
with $\Mstars \la 10^{11} \Msun$ (e.g.\ \citealt{Cat08,Str08,Fon09}).

Given the varying results on galaxy evolutions both observationally and 
theoretically, it would be of great value to constrain galaxy evolutions
through other independent methods based on different data sets.
The statistical analysis of strong lensing galaxies in conjunction with 
well-determined local velocity dispersion functions (VDFs) of galaxies provides
 such an independent method. The VDFs represent the statistical properties of 
the  dynamics of galaxies that have been modified by baryonic physics from the
CDM haloes. The velocities of  particles and stars are 
boosted and the density profiles become 
steeper. These dynamical features can be seen
by comparing the dynamical properties of $N$-body simulated haloes and the 
observationally-derived kinematical and dynamical properties of visible
galaxies. N-body simulations show that pure CDM haloes follow
NFW (or NFW-like) density profiles (e.g.\ \citealt{NFW97,Moo99,JS02,Nav04}) 
and hence rising velocity profiles in the
inner parts whereas the kinematical and dynamical properties of galaxies,
derived from such methods as stellar dynamics (e.g.\ \citealt{Rix97,Cap06}), 
rotation curves (e.g.\ \citealt{Rub85,Per96,Sal07}), gravitational lensing 
(e.g.\ \citealt{RK05,Koo06,Gav07}), show that 
baryon settled galaxies follow isothermal
(or isothermal-like) density profiles implying that velocities are boosted and
mass distributions become more concentrated in the inner regions. An important
observable quantity that characterises the dynamical state of the galaxy is 
the line-of-sight velocity dispersion $\sigma$ especially for rotation-free
early-type galaxies. For late-type galaxies, rotations are important in the
disk. Nevertheless, the bulge and the surrounding halo of the late-type galaxy
may be regarded as an ellipsoidal system that is similar to early-type galaxies.
Furthermore, given that most late-type systems are isothermal(-like) as
inferred from nearly flat rotation curves, we may relate the circular rotation
speed in the disk $v_c$ to the velocity dispersion in the inner halo
 via $v_c = \sqrt{2}\sigma$. 

All three statistical functions (i.e.\ LF, SMF, and VDF) 
are the consequences of baryonic physics from pure CDM haloes. However, unlike 
the other functions the VDF has only to do with the dynamical effects of 
baryonic physics separated from many other complex effects involving radiative
processes.   Furthermore, given that galaxy mass profiles are likely to be 
isothermal(-like) up to significant fractions of the dark halo virial radii as
stressed above, we expect a good correlation between the central velocity 
dispersion and the halo virial mass.
Hence the velocity dispersion function (VDF) will be another independent and
powerful constraint on galaxy formation and evolution. 
One advantage of the VDF as a cosmological probe is that it 
can test the underlying $\LCDM$ structure formation theory more robustly
because it is linked more intimately to the theoretical halo mass function.

In this paper we use a range of local VDFs. The local VDFs include not only the
 SDSS VDF of early-type galaxies measured by \citet{Cho07} but also the 
type-specific and total VDFs obtained through a Monte Carlo method based on  
the galaxy  LFs from the SDSS and intrinsic correlations 
between luminosity and velocity dispersion or circular rotation speed. 
We then constrain the evolutions of the VDFs 
through the statistical analysis of 
galactic-scale strong gravitational lensing. Strong lensing statistics
is a powerful means to constrain the evolution of the VDF because strong 
lensing probability is proportional to comoving number density times $\sigma^4$
 (the image separation is  proportional to $\sigma^2$)   
and lensing galaxies are distributed over a
range of redshifts (up to $z \ga 1$) and virtually over all area of sky.
The latter property is particularly important since it means that 
unlike deep galaxy surveys strong lensing data  are free from cosmic variance.
For strong lensing statistics we use a total of 30 lenses from the 
radio-selected Cosmic Lens All-Sky Survey (CLASS; \citealt{Mye03,Bro03}) and
PMN-NVSS Extragalactic Lens Survey  (PANELS; \citealt{Win01}) and the 
optically-selected Hubble Space Telescope Snapshot survey (\citealt{Mao93}), 
as these surveys have well-defined selection functions  and cover about 
the same image separation range (i.e., $\ga 0''.3$) and similar
redshift ranges for source quasars.\footnote{SDSS quasar lens search 
(\citealt{Ina08}) is limited to image separation range of $\ge 1''$ and 
source redshift range of $0.6 < z < 2.2$.} However, only the CLASS
statistical sample of 13 lenses (\citealt{Bro03,Cha03}) and the Snapshot
sample of 4 lenses are used for the statistics of absolute lensing while the 
rest of lenses are used only for the distributions of image separations (and 
lens redshifts if available).

This paper is organised as follows. In \S2 we derive through a Monte Carlo 
method the local VDFs for both the type-specific galaxy populations and the 
total population. In \S3 we describe the strong lensing data
and the theoretical model for statistical analysis of strong lensing.
In \S4 we constrain (non-evolving) intermediate redshift VDFs using strong 
lensing data and then compare the intermediate redshift VDFs with local VDFs. 
This simple analysis captures the trends of the evolutions.
Next we study in detail the evolution of the VDF through a parametric approach 
(\S 5). The results that come out are striking. The number density evolution
as a function of $\sigma$ is differential: the number density of galaxies 
of typical velocity dispersion and lower evolves little since $z=1$ but that 
of larger velocity dispersion evolves significantly showing 
``velocity upsizing (hierarchical)'' behaviour of galaxy evolution. 
In the discussion section we first consider possible systematic errors (\S6.1)
and then compare with previous results of strong lensing on galaxy 
evolutions (\S6.2) and the evolutionary behaviours of the DMF and the SMF from 
the literature (\S6.3). A comprehensive analysis of the coevolution of the DMF,
the VDF and the SMF is carried out in a companion paper (\citealt{Cha10}).
We give the conclusions in \S7. Throughout we assume a $\LCDM$ cosmology with 
$(\Omega_{{\rm m}0},\Omega_{{\Lambda}0})=(0.25,0.75)$ and
$H_0 =100h \kms$~Mpc$^{-1}$ consistent with the WMAP 5 year data
(\citealt{Dun09}). When parameter $h$ does not appear explicitly, 
 $h=0.7$ is assumed.

\section{Velocity dispersion functions of galaxies at $z \approx 0$}

\subsection{Statistical functions of galaxies}

The Schechter luminosity function (LF) $\phi_{\rm L}$,  the differential
comoving number density as a function of luminosity, is given by
\beq
dn = \phi_{\rm L} (L) dL = \phi_{\rm L}^*
  \left(\frac{L}{L^{*}}\right)^{\alpha_{\rm L}}
   \exp\left(-\frac{L}{L^{*}}\right)
    \frac{dL}{L^{*}}.
\label{LF}
\eeq
It has been measured extensively at various wavebands and recent observations 
have produced reliable results at optical (e.g.\ \citealt{Nor02,Bla03,Cro05})
 and infrared (e.g.\ \citealt{Col01,Koc01}) wavebands owing 
particularly to large surveys such as the Sloan Digital Sky Survey (SDSS), 
the Two Degree Field Galaxy Redshift Survey (2dF) and the Two Micron All Sky
Survey (2MASS). 

\begin{figure*}
\begin{center}
\setlength{\unitlength}{1cm}
\begin{picture}(14,7)(0,0)
\put(-2.,11.){\includegraphics{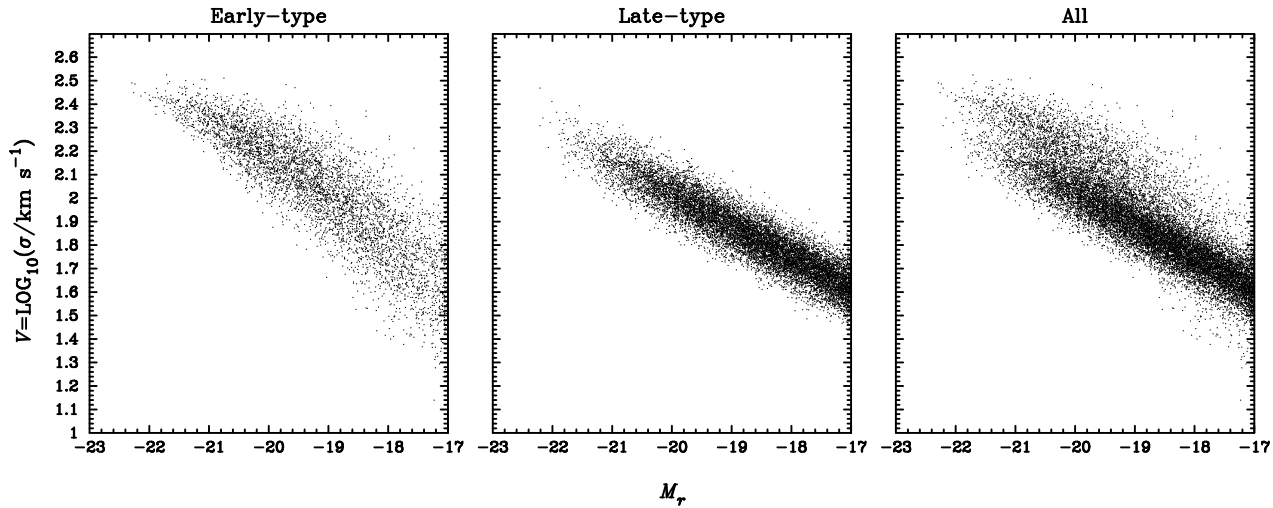}}
\end{picture}
\caption{
Monte Carlo realised magnitude-velocity dispersion ($M_r - \sigma$) relations. 
Only 5\% of the generated galaxies are shown.
The magnitude is the SDSS $r$-band absolute magnitude. 
The relation for the early-type population is
from the \citet{Cho07} measured relation while that for the late-type 
population is inferred from the relation between $M_r$ and the rotational 
circular speed by \citet{Piz07} assuming a singular isothermal sphere
for the total mass distribution of a galaxy. The relation for all galaxies
is the combination of the early-type and the late-type relations.
}
\label{MVplane}
\end{center}
\end{figure*}

Given the observed power-law correlations between luminosity and 
internal velocities of galaxies such as the Faber-Jackson (\citealt{FJ76})
 relation and the Tully-Fisher (\citealt{TF77}) relation, one may expect 
a modified form of the Schechter function for a velocity function. 
Indeed, \citet{She03} find that the distribution of the velocity 
dispersions of early-type galaxies from the SDSS is well fitted by 
the VDF $\phi$ of the form
\beq
dn = \phi (\sigma) d\sigma = \phi^*
  \left(\frac{\sigma}{\sigma^*}\right)^{\alpha}
   \exp\left[-\left(\frac{\sigma}{\sigma^*}\right)^{\beta}\right]
   \frac{\beta}{\Gamma(\alpha/\beta)} \frac{d\sigma}{\sigma},
\label{VDF}
\eeq
or equivalently
\beq
 \phi (V) dV = \phi^* 10^{\alpha(V-V^*)} 
  \exp\left[-10^{\beta (V-V^*)}\right]
   \frac{\beta \ln 10 }{\Gamma(\alpha/\beta)} dV,
\label{VDFlog}
\eeq 
where $V$ is the logarithmic  variable given by
\beq 
 V = \log_{10} \sigma.
\label{V}
\eeq
Notice that equation~(\ref{VDF}) along with equation~(\ref{LF}) imply the
following correlations:
\beq
 \alpha_{\rm L}= \alpha/\beta -1 \hspace{1em} {\rm and} \hspace{1em}
 \phi_{\rm L}^* =\phi^*/ \Gamma(\alpha/\beta),
\label{corr}
\eeq
 where $\phi^*$ is the integrated comoving number density of galaxies, and
 an `effective' power-law correlation between luminosity ($L$) and 
velocity dispersion ($\sigma$) of
\beq
\frac{L}{L^*} =   \left(\frac{\sigma}{\sigma^*}\right)^{\beta}.
\label{LVD}
 \eeq
However, directly measured relations of $L=L(\sigma)$ and $\sigma=\sigma(L)$
give different values for the power-law slope $\beta$ compared with that
in equation~(\ref{LVD}) because the scatter in the correlation acts in 
different ways for the different fits (\citealt{She03}). Furthermore, the
value of $\beta$ varies as a function of luminosity as can be seen in a
magnitude-velocity dispersion plane of galaxies 
[see Fig.~\ref{MVplane} or {Fig.}~4b of \citet{Cho07}].
This is why it is not reliable to turn a LF of galaxies to a VDF simply using
equation~(\ref{corr}) and a (mean) Faber-Jackson or Tully-Fisher relation 
although this was often practised in the past. 
One must fully take into account the
intrinsic scatter in the magnitude-velocity dispersion plane.
This is done below.

\subsection{VDF of early-type galaxies}

\citet{Cho07} have measured
the VDF for local early-type galaxies  based on a large sample of
galaxies from the SDSS DR5 data set, employing a galaxy classification scheme
that closely matches visual classifications (\citealt{PC05}). The early-type 
VDF measured by \citet{Cho07} will be referred to as `E0' VDF and can be found
in Table~\ref{GFs}. Notice, however, 
that galaxy number counts start to become incomplete at low velocity
dispersions for a given galaxy sample used by \citet{Cho07} because the 
sample is not limited by velocity dispersion but by absolute magnitude. 
  \citet{Cho07} discuss this kind of problem for $\sigma < 150 \kms$ and
obtain galaxy counts down to $\sigma \approx 70 \kms$ using a series of
samples of different volume and magnitude limits. The measurement of the LF
is free of this problem at low luminosities down to the absolute magnitude
 limit of a given sample. Hence galaxy counts are more reliable at low 
luminosities than low velocity dispersions. 

It is therefore desirable to use galaxy counts at low luminosities in deriving 
the VDF for the full range of velocity dispersions. 
We use a Monte Carlo method as follows. We generate galaxies within a comoving
volume  brighter than an absolute magnitude limit  according to
a measured luminosity function.  
The next step is to assign a velocity dispersion to each galaxy of a given
absolute magnitude. To do this we use the 2-dimensional distribution of galaxies
in the absolute magnitude-velocity dispersion ($M-\sigma$) plane.
We assume that the most likely value of $V \equiv \log_{10} \sigma$ varies
as a function of $M_r$ (absolute magnitude in SDSS $r$ band) according to
 the \citet{Cho07} {Fig.}~4b.
We assume that $V$ follows a normal distribution for a given $M_r$ with
a varying dispersion of
\beq
s(M_r)=a(M_r - M_r^*)+s^*,
\label{disp}
\eeq
as suggested by the \citet{Cho07} {Fig.}~4b.
Once all mock galaxies are assigned velocity dispersions, we finally derive 
the VDF by counting mock galaxies as a function of $V$.

\begin{figure}
\begin{center}
\setlength{\unitlength}{1cm}
\begin{picture}(8,11)(0,0)
\put(-1.2,-0.7){\includegraphics{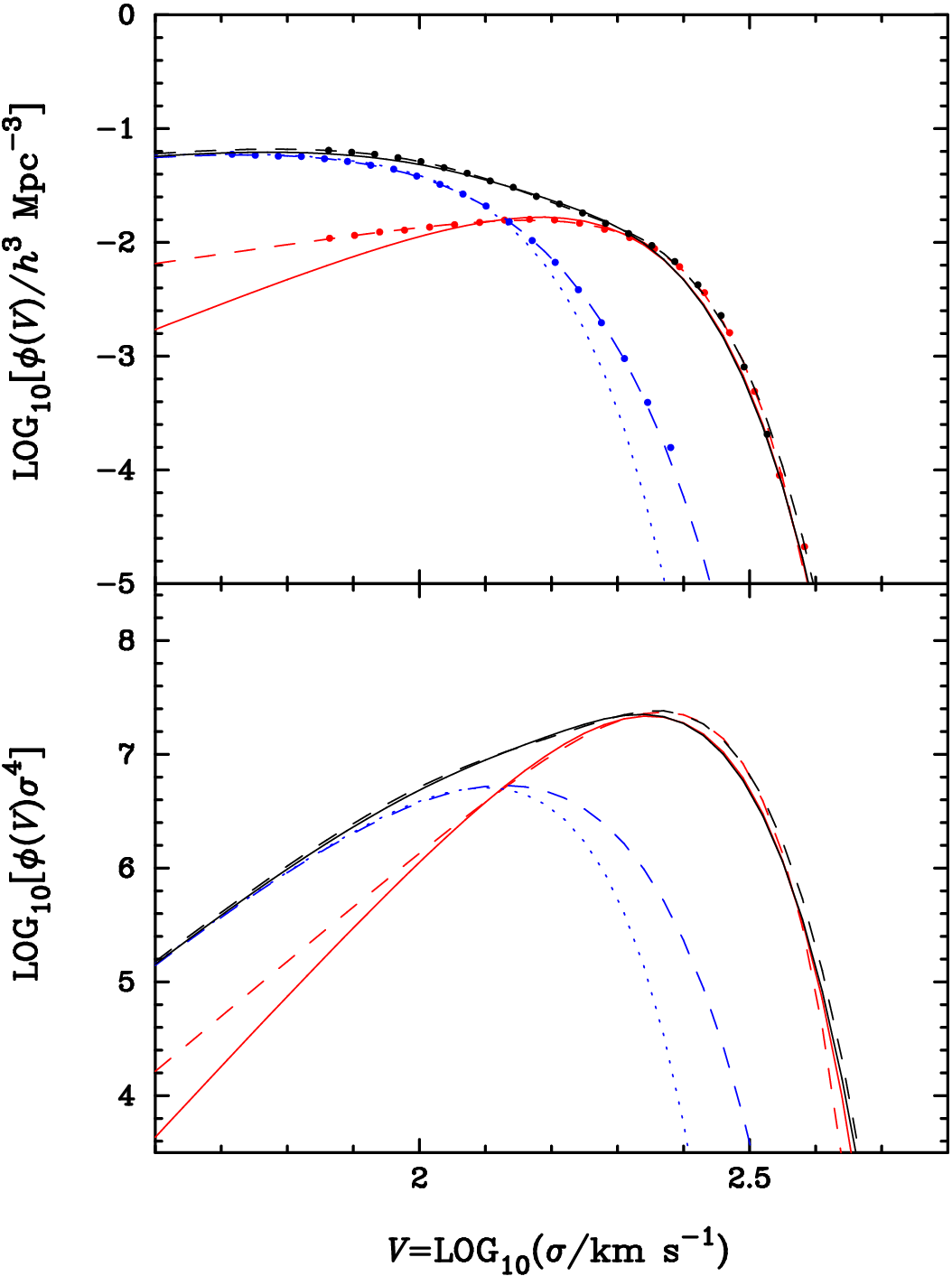}}
\end{picture}
\caption{
({\it Upper panel}) The local velocity dispersion functions of galaxies based 
on SDSS data (see also Table~\ref{GFs}). The red, blue and black curves 
represent respectively the VDFs for the early-type, the late-type and all 
populations. The dashed curves are Monte Carlo realised VDFs. The red solid 
curve is the early-type E0 VDF measured by \citet{Cho07}. The black solid
curve is the total VDF resulting from the combination of the E0 VDF and the 
Monte Carlo realised late-type VDF. The blue dotted curve is the late-type VDF
that would be obtained ignoring the intrinsic scatter in the magnitude-velocity
relation. 
({\it Lower panel}) The behaviours of $\phi(V) \times \sigma^4$ where $\phi(V)$
is the differential number density of galaxies and $\sigma^4$ is proportional
to the lensing cross section of each galaxy (assuming the isothermal galaxy 
profile). These curves show the relative lensing efficiencies of the early-type
and the late-type galaxy populations as a function of velocity dispersion.
}
\label{VDFs}
\end{center}
\end{figure}

We consider the SDSS early-type LF by \citet{Cho07}. The \citet{Cho07} LF is
the optimal choice for the following two reasons. First,
\citet{Cho07} classify galaxies using a classification method that
 matches closely visual classification (\citealt{PC05}). 
Second, the \citet{Cho07} LF is based on the same SDSS magnitude system that 
is used for the magnitude-velocity dispersion relation, so that any error
arising from photometric transformations (e.g., from the 2dF magnitude) is 
avoided. For this LF we derive the early-type VDF
using the Monte Carlo method described above adopting 
$M_r^*-5 \log_{10} h =-20.23$, $s^*=0.085$ and $a=0.025$ where
$s^*$ and $a$ are chosen so that the simulated $M_r - V$ relation mimics that
observed by \citet{Cho07} with a mean dispersion of $\approx 0.11$ for
$-22 \la M_r \la -17.5$. The simulation
comoving volume is $10^7$~$h^{-3}$~Mpc$^3$ and the faint magnitude limit is 
$M_r-5 \log_{10} h =-16$. This simulation generates more than
100,000 early-type galaxies.
The simulated $M_r - V$ relation for the \citet{Cho07} early-type LF can be 
found in {Fig.}~\ref{MVplane}. Notice that this simulated plane is very similar
to the \citet{Cho07} Fig.~4b. The derived early-type VDF will be referred to as
 `E1' VDF and can be found in {Fig.}~\ref{VDFs} and table~\ref{GFs}. 
Notice that the VDF derived from the
\citet{Cho07} early-type LF closely matches the \citet{Cho07} measured VDF
 for $\sigma \ga 150 \kms$ but shows significant difference at lower $\sigma$.
This could mirror the incompleteness in galaxy counts at low velocity 
dispersions by \citet{Cho07}. Because of this possibility we consider both the
\citet{Cho07} measured VDF and the Monte Carlo realised VDF.

\begin{table*}
\caption{Local statistical functions of galaxies based on SDSS data
\label{GFs}}
\begin{tabular}{cccccccccl}
\hline
 Type & ID & $\alpha$  & $\beta$  & $\alpha_{\rm L}$ & 
   $\phi^*$ &  $\phi_{\rm L}^*$ &  $M_{r}^* - 5 \log_{10} h$
    & $\sigma^*$  & Reference  \\
   & & &  &  & ($h^3$ ${\rm Mpc}^{-3}$) &  ($h^3$ ${\rm Mpc}^{-3}$) &   
      &  ($\kms$)  &   \\ 
\hline
\multicolumn{10}{c}{Early-type} \\
\hline
  VDF & E0 & 2.32& 2.67& $-0.13$ & 0.008 & 0.0073& $\cdots$ & 161.&
   \citet{Cho07}\\
   LF & $\cdots$  & $\cdots$ & $\cdots$ & $-0.527$ & $\cdots$ & 0.0071 & 
  $-20.23$ &  $\cdots$     & \citet{Cho07}\\
  VDF& E1 & 0.85 & 3.72 & $-0.77$& 0.013 & 0.0033& $\cdots$& 217.& This work\\
\hline
\multicolumn{10}{c}{Late-type} \\
\hline
 LF & $\cdots$& $\cdots$ & $\cdots$ & $-0.897$& $\cdots$ & 0.012 & 
  $-20.22^a$ & $\cdots$     & \citet{Cho07} \\
 VDF& L1 & 0.69 & 2.10 & $-0.67$& 0.066& 0.024& $\cdots$& 91.5& This work \\
\hline
\multicolumn{10}{c}{All} \\
        &   &   &   & $\alpha'$ &   & $\varepsilon$ &  &   &    \\ 
\hline
 VDF & A0  & 0.69 & 2.01 & 6.61 & 0.074 & 0.044 & 
  $\cdots$ & 100.   & This work \\ 
    VDF & A1  & 0.69 & 2.05 & 7.69 & 0.078 & 0.035 & $\cdots$ & 102.
           & This work \\ 
\hline
\end{tabular}

Notes. 
 $^a$  An internal extinction of $A_r \approx 0.1$ 
 [extrapolated from {Fig.}~11 of \citet{Piz07}] is assumed and corrected for
 the \citet{Cho07} $b/a > 0.6$ late-type galaxies.   

\end{table*}

\subsection{VDF of late-type galaxies}

There does not yet exist any kind of directly measured velocity functions for 
the late-type population of galaxies. For the late-type population, the internal
motions in the optical regions of galaxies are observed as rotations 
since the (random) motions of dark matter particles  are not observed. 
Unfortunately, it is not an easy observational task to measure the rotation of
a galaxy and consequently rotations of galaxies have not been measured for
all galaxies in a volume or magnitude limited galaxy sample. 
However, one may obtain a velocity function for the late-type population 
from the LF through the Tully-Fisher relation between rotational
speed and luminosity that is usually a tighter correlation than the
Faber-Jackson relation of the early-type population. For example, \citet{Gon00},
 \citet{KW01}, \citet{Cha03} and \citet{She03} inferred  circular velocity 
functions (CVFs) of late-type galaxies combining various late-type LFs and 
Tully-Fisher relations. It is also possible to convert the CVF to the VDF
via $\sigma=\vc /\sqrt{2}$ assuming the singular isothermal sphere model for
the total mass distribution of a galaxy.\footnote{Dark matter haloes are, 
in general, triaxial in shape and have density profiles that deviate from the
isothermal profile. Nevertheless, the deviations of shapes and profiles from
the SIS may have cancelling effects statistically so that the averaged 
property of a galaxy might not deviate too much from the SIS.} 
In this work we essentially follow this approach to obtain  estimates
 of the VDF of late-type galaxies. However, we fully take into account the
intrinsic scatter in the Tully-Fisher relation using the Monte Carlo method
described in \S2.2.

The Tully-Fisher relation has been published  by numerous authors. 
The relation is usually expressed as $L/L^* = (\vc / \vcstar )^\beta $ or
its inverse relation  $\vc / \vcstar = (L/L^*)^{1/\beta}$.
The slope $\beta$ ranges from $\approx 2.5$ - $ 3.5$ in most of 
the published results in optical and infrared bands (See \citealt{Piz07} and
references therein). 
\citet{Piz07} derive the relation $L=L(\vc )$ and its inverse relation 
$\vc = \vc (L)$ for 162  SDSS galaxies in the SDSS g, r, i and z bands.
The galaxies used by \citet{Piz07} have apparent minor-to-major axis ratio 
$b/a \le 0.6$ so that internal extinctions are measured and corrected for
reliably.
The measurements of the TF relation and its inverse relation
give different slopes because of the scatter (as for the FJ relation).
Notice that the inverse relation $\vc / \vcstar = (L/L^*)^{1/\beta}$ is
more applicable in deriving a velocity function from the LF because
the velocity function is estimated from the LF. 
For the SDSS $r$-band \citet{Piz07} find 
$\log_{10} (\vc) = (-0.135 \pm 0.006) (M_r -M_r^*) + (2.210 \pm 0.006) $
with $M_r^* -5 \log_{10} h = -20.332$ and a dispersion of $0.063\pm 0.005$.  
The relation between $V(=\log_{10}\sigma)$ and $M_r$ then follows from
$\log_{10} (\vc)=V+ \log_{10} \sqrt{2} $ for the SIS model of the galaxy.

Using the Monte Carlo method based on the above \citet{Piz07} $M_r - V$ 
relation the VDF of late-type galaxies is derived for the late-type LF by 
\citet{Cho07}. 
As in \S2.2 the simulation comoving volume is $10^7$~$h^{-3}$~Mpc$^3$ and
 the faint magnitude limit is $M_r-5 \log_{10} h =-16$. 
By this simulation more than
300,000 late-type galaxies are generated.
The derived VDF will be referred to as `L1' VDF and
can be found in {Fig.}~\ref{VDFs} and Table~\ref{GFs}.
{Fig.}~\ref{VDFs} also shows a VDF derived assuming no intrinsic scatter 
for the \citet{Cho07} late-type LF to demonstrate the effect of the intrinsic 
scatter in the Tully-Fisher relation.
The simulated $M_r - V$ relation for the \citet{Cho07} late-type LF can be 
found in {Fig.}~\ref{MVplane}.

\subsection{VDF of all galaxies}

In \S2.2 and \S2.3 the VDFs are derived for morphologically-typed galaxies.
One may well expect that a well-defined VDF exist for the entire population
of galaxies. Here we consider this possibility. The total VDF is motivated
in several respects. First of all, the total VDF can be compared with 
theoretical predictions more straightforwardly than type-specific VDFs 
just as is the total LF because galaxy classifications are not required. 
Secondly, the evolution of the total VDF is of interest in its own right 
as it represents the collective evolution of all galaxies including 
morphological transformations through merging. Thirdly, strong lensing
computation is simplified if the total VDF is used because one does not
require knowledge of the lensing galaxy type which is not well determined
by observations for some systems. Lastly, the observed evolutions of the total 
LF/SMF may be used for or compared with the evolution of the total VDF.

It is straightforward to derive the total VDF using the Monte Carlo method 
used above. Early-type and late-type galaxies are separately generated and then
combined in the $M_r - V$ plane. For the comoving volume of 
$10^7$~$h^{-3}$~Mpc$^3$ and the faint magnitude limit of $M_r-5 \log_{10} h =-16$
over 400,000 velocity dispersions of galaxies are  used to derive the
total VDF. Again we consider the LFs by \citet{Cho07}. We also consider
the early-type VDF directly measured by \citet{Cho07}.
 The derived total VDFs can be found in {Fig.}~\ref{VDFs}.
The `A0' VDF corresponds to the combination of the E0 VDF and the L1 VDF
while the `A1' VDF corresponds to that of the E1 and the L1. 
 Notice that the simulated data points for the total population
of galaxies are not fitted well by the VDF of equation~(\ref{VDF}) in contrast 
to the morphologically-typed galaxy populations. This might be the true nature
of the total VDF. However, it could be due to the errors in the adopted 
 correlations between luminosity and velocity especially at low 
velocities.\footnote{Here it is interesting to note that the early-type and
the late-type populations are well fitted by Schechter luminosity functions
both separately and in combination although they are well separated 
photometrically.}  

In order to fit  successfully the total VDF from the present data we consider
the following function that includes a correction term for high velocity 
dispersions, 
\begin{eqnarray}
 \phi_{\rm tot} (\sigma) d\sigma & = & \phi^* \left[
 (1-\varepsilon) \left(\frac{\sigma}{\sigma^*}\right)^{\alpha}
 + \varepsilon \frac{\Gamma(\alpha/\beta)}{\Gamma(\alpha'/\beta)}
    \left(\frac{\sigma}{\sigma^*}\right)^{\alpha'}
 \right]  \nonumber \\
 & & \times  \exp\left[-\left(\frac{\sigma}{\sigma^*}\right)^{\beta}\right]
   \frac{\beta}{\Gamma(\alpha/\beta)} \frac{d\sigma}{\sigma},
\label{VDFtot}
\end{eqnarray}
where $\phi^*$ is the total number density and $\varepsilon$ represents the 
fraction of the number density due to the second (correction) term.
{Fig.}~\ref{VDFs} shows that the total VDFs are well fitted by 
equation~(\ref{VDFtot}).  For the fitting process there arises 
a weak degeneracy between parameters $\alpha$ and $\varepsilon$. 
To break the degeneracy we fix the value of $\alpha$ 
to that for the late-type VDF since late-type galaxies dominate in the low 
velocity dispersion limit. The fitted parameters of the total VDFs can be found
in Table~\ref{GFs}.

\section{Strong lensing statistics: data and model}

\subsection{Data}

For the purpose of this work we need galactic-scale strong lens samples that
satisfy well-defined observational selection criteria. Such statistical lens
samples that have been published so far include those from  the radio-selected 
CLASS and PANELS surveys (\citealt{Mye03,Bro03,Win01}) and the 
optically-selected HST Snapshot (\citealt{Mao93}) and SDSS 
Quasar Lens Search (QLS: \citealt{Ina08}) surveys. However, the SDSS QLS lens 
systems are limited to image separations greater than 1 arc second while
the other surveys include subarcsecond systems down to $\approx 0''.3$. 
Moreover, the SDSS QLS lens systems are from relatively low-redshift 
($0.6 < z < 2.2$) sources while the other surveys include high-redshift
sources.   For these reasons we use only the HST Snapshot and the CLASS/PANELS
 lens samples to study the evolutions of the VDFs. 
The HST Snapshot data include 4
lens systems out of 506 quasars with $z > 1$. The HST Snapshot lens systems
are summarised in Table~\ref{lens} and the data for the entire sources can be 
found in \citet{Mao93}.  We use the selection function for the Snapshot survey 
described by \citet{MR93} and the source number-magnitude relation described 
by \citet{Koc96} to calculate magnification biases and cross sections. 
The CLASS data include 22 lens systems out of 16521 radio sources. However,
only a subsample of 8958 sources including 13 lens systems, referred to
as the CLASS statistical sample, satisfies 
well-defined selection criteria (see \citealt{Cha03,Bro03}) so as to allow
reliable calculation of cross sections and magnification biases. 
We use the CLASS selection function and source properties described by 
\citet{Cha03} except that we use the number-flux density relation slightly 
updated by \citet{McK07a}. 
The rest of 9 lens systems from the CLASS and additional 4 lens 
systems from the PANELS (the southern sky counterpart of the CLASS)
can be used for relative lensing probabilities (see \citealt{Cha05}).
The CLASS and PANELS lens systems can be found in Table~\ref{lens}.

\begin{table*}
\caption{Summary of Strongly-lensed Sources.
The properties of the strongly lensed systems from the 
Snapshot optical survey and the CLASS (\citealt{Bro03}) and PANELS radio
surveys are revised from \citet{Koc96} and \citet{Cha03, Cha05}.
 The first 13 radio sources (B0218+357 through B2319+051) 
are members of the well-defined CLASS statistical sample of 8958 radio sources 
(\citealt{Bro03, Cha03}). The CLASS statistical sample along with the 
Snapshot sample are used fully for all their observed properties while the rest
of the sources are used only for their relative image separation probabilities.
Notations are the following: $m_B$ - $B$ magnitude; 
$f_5$ - flux density at 5~GHz; $z_s$ - source redshift; $z_l$ - lens redshift;
$\Delta\theta$ - the maximum separation between any pair of the images; 
$\twoRi$ - twice the average separation of the images from the primary lensing
galaxy (or the image centre if the lens is not observed);  
$N_{\rm im}$ - number of images; E/L - early/late-type galaxy; 
CL - lensing cluster/group; G2 - secondary (satellite) lensing galaxy.
References are the following: 1 - the CASTLES website 
({\mbox{http://cfa-www.harvard.edu/castles/}});  2 - \citet{Cha03,Cha05}; 
3 - \citet{Koc96}; 4 - \citet{Sie98}; 5 - \citet{Yor05}; 6 - \citet{Big03}; 
 7 - \citet{Rus01}; 8 - \citet{SB03}; 9 - \citet{Coh01}; 10 - \citet{McK07b};
 11 - \citet{Cha01}; 12 - \citet{Phi00}; 13 - \citet{Win03}; 14 - \citet{Leh00};
 15 - \citet{Kin97}; 16 - \citet{Mor08}
\label{lens}}
\begin{tabular}{lllllccllll}
\hline
Source &  Survey & $m_{\rm B}$/ & $z_s$ & $z_l$  &  $\Delta\theta$ 
  & $\twoRi$   & $N_{\rm im}$   & Lens Type & Comments & References \\
   &   &  $f_{5}$ [mJy]  &   &  & [$''$]   & [$''$] &  &   &  &  \\    
\hline
Q0142-100 & Snapshot & 16.8  & 2.72 & 0.49 & 2.23& 2.24 & 2 & E & & 1, 3 \\
PG1115+080 & Snapshot & 16.1  & 1.72 & 0.31& 2.43& 2.32 & 4 & E & Group& 1, 3\\
Q1208+1011 & Snapshot & 17.9  & 3.80 & 1.13? & 0.48& 0.48 & 2 & ? & & 1, 3, 4 \\
H1413+117 & Snapshot & 17.0  & 2.56 & ---  & 1.10& 1.24 & 4 & ?(E)$^a$ & 
           & 1, 3 \\
   &   &   &   &  &  &  &   &  \\    
B0218+357  & CLASS & 1480.  & 0.96 & 0.68 & 0.33& 0.34 & 2 & L & & 1, 2 \\
B0445+123 & CLASS  & 50.   & --- & 0.56 & 1.32& 1.35 & 2 & E & & 1, 2  \\
B0631+519 & CLASS  & 88.   & --- & 0.62 & 1.16& 1.17 & 2 & E & & 2, 5  \\
B0712+472 & CLASS  & 30.   & 1.34 & 0.41 & 1.27& 1.46 & 4 & E & Group& 1, 2 \\
B0850+054  & CLASS & 68.   &  ---   & 0.59 & 0.68& 0.68 & 2& L & & 2, 6  \\
B1152+199 & CLASS & 76.  & 1.02 & 0.44 & 1.56& 1.59 & 2 & 
                      2Gs [?(E)$^a$+?] &   G2 ignored &  1, 2  \\
B1359+154 & CLASS & 66. & 3.24 & $\approx 1$ & 1.67& 1.71  & 6 & 3Gs (E+?+?) & 
          Group &  1, 2, 7   \\
B1422+231 & CLASS  & 548.  & 3.62 & 0.34 & 1.24& 1.68& 4 & E & Group &  1, 2 \\
B1608+656  & CLASS &  88.  & 1.39 & 0.64 & 2.09& 2.34& 4 & 2Gs (E+L) & 
         Group & 2, 8  \\
B1933+503 & CLASS  & 63.   & 2.62 & 0.76 & 1.16& 1.02 & 4+4+2 & E &  & 2, 9 \\
B2045+265 & CLASS & 55.  & ?$^b$ & 0.87 & 1.91& 2.34& 4& 2Gs (E+?) &  & 2, 10 \\
B2114+022 & CLASS& 224.  & ---& 0.32/0.59& 2.56& 2.62& 2 or 4 & 2Gs (E+E) & 
          Group & 2, 11  \\
B2319+051  & CLASS  & 76.  & --- & 0.62 & 1.36& 1.36 & 2 & E & Group& 1, 2 \\
   &   &   &  &  &   &   &   &  &   \\    
B0128+437 & CLASS & --- & 3.12 & 1.15 & 0.55& 0.46 & 4 & ? &  & 2, 12  \\
J0134$-$0931& PANELS&---& 2.23 &0.76 & 0.68& 0.81 & 5+2& 2Gs (L?+L?)& 
       & 1, 2, 13  \\
B0414+054  & CLASS & ---& 2.64 & 0.96 & 2.03& 2.40 & 4  & E & & 1, 2  \\
B0739+366 & CLASS & ---   & ---   &  ---  & 0.53& 0.53 & 2  & ? & & 1, 2  \\
B1030+074 & CLASS & --- & 1.54 & 0.60& 1.62& 1.63 & 2 & 2Gs (E+?) 
        & G2 ignored  & 1, 2, 14  \\
B1127+385 & CLASS  & ---  &   ---  &  ---  & 0.71& 0.74 & 2  & L &  & 1, 2  \\
B1555+375 & CLASS  & ---  &  ---   &  ---   & 0.41& 0.47 & 4  & ? &  & 1, 2  \\
B1600+434 & CLASS  & ---  & 1.59 & 0.41 & 1.38& 1.40 & 2  & L &  Group& 1, 2 \\
J1632$-$0033 & PANELS & --- & 3.42  &  1?  & 1.47& 1.47& 2 & E &  & 1, 2  \\
J1838$-$3427 & PANELS& ---  & 2.78  & 0.36 & 0.99& 0.99 & 2  & E & & 1, 2 \\
B1938+666 & CLASS& ---& $\ga 1.8$ & 0.88 & 0.91& 0.85 & 4+2+R & E & & 1, 2, 15\\
J2004$-$1349 & PANELS& ---  & --- &  ---  & 1.13& 1.18 &  2  & L &  & 1, 2  \\
B2108+213 & CLASS& ---& --- & 0.37  & 4.57& 4.57 & 2 & 2Gs (E+E) & 
 Group  & 1, 2, 16  \\
\hline
\end{tabular}

$^a$ Assumed early-type for their large inferred velocity dispersions of
 $\sigma \approx 270\kms$ for H1413+117 and
 $\sigma \approx 230\kms$ for B1152+199.

$^b$ \citet{Fas99} suggests $z_s =1.28$ based only on a single broad emission 
line. This value for $z_s$ implies an uncomfortably large SIS velocity 
dispersion of $\sigma \approx 384 \kms$ for G1. 
However, \citet{Ham05} reports a measured
central stellar velocity dispersion of $\sigma=213\pm 23 \kms$.
 An alternative value for  $z_s$ would be 4.3 assuming the emission line is 
a Lyman alpha line. We take the latter value for this work.

\end{table*}

Notice that in Table~\ref{lens} both the image separation $\Delta\theta$ and 
twice the average radial separation $\twoRi$ are given for each lens system. 
$\Delta\theta$ is the maximum possible
separation amongst the image components. This is the value adopted in
previous studies (e.g.\ \citealt{Cha03,Cha05}). 
 $\twoRi$ is two times the average of the separations of 
the image components from the primary lensing galaxy position or 
the image centre if the lensing galaxy position is not known. 
If the lens were an SIS, the two values would be the same.
However, for many systems the two values are somewhat different, 
$\twoRi$ usually being somewhat larger than $\Delta\theta$. 
The systems with relatively large differences tend to be 
quadruply imaged systems with highly asymmetrical morphologies. 
The average values for all 30 lens systems are $\Delta\theta=1''.38$ and 
$\twoRi=1''.45$ respectively while those for 9 lens systems that are only 
quadruply imaged are $\Delta\theta=1''.45$ and $\twoRi=1''.63$ respectively. 
If we exclude 6 multiple-lens systems, 
the average separations for the rest 24 systems
 are $\Delta\theta=1''.17$ and $\twoRi=1''.21$ respectively.  
Although the differences between $\Delta\theta$ and $\twoRi$ are not large, 
we consider the two possibilities in order to be as precise as possible. 
The question is then which of the two choices would correspond
 more accurately to the velocity dispersion of the primary lensing galaxy. 
Because the velocity dispersion is directly related to the Einstein radius
$R_{E}$ (i.e.\ the critical radius) via a lens model 
(see, e.g., \citealt{Cha03}), the answer may be found by comparing
 $\Delta\theta$ and $\twoRi$ with $2 R_{E}$. 
The Einstein radius $R_{E}$ for a lens can only be reliably determined from 
a detailed modelling of the system. This is particularly true for highly 
asymmetrical systems with large shears. \citet{CK05} model some of highly
asymmetrical quadruple lens systems through a lens model with the isothermal 
radial profile and general angular structures. The highly asymmetrical systems 
that are analysed by \citet{CK05} include B0712+472, B1422+231 and B2045+265.
As shown in Table~\ref{lens}, for these systems the values for $\Delta\theta$
and $\twoRi$ are quite different. If we take the Einstein radii found by 
\citet{CK05} for these systems, then $2 R_{E}$ are between $\Delta\theta$
and $\twoRi$ but much closer to $\twoRi$.
This means that $\twoRi$ is probably the better choice for this work. Hence 
the results from this work are based on  $\twoRi$ rather than $\Delta\theta$.

\begin{table*}
\caption{Multiple-lens systems
\label{multi}}
\begin{tabular}{lcccccll}
\hline
System& G1's ER& $2R_{\rm E}^{(1)}$& G2's ER & G3's ER& $\Ri - R_{\rm E}^{(1)}$
 & Comments & Reference \\
  & $R_{\rm E}^{(1)}$ &   & $R_{\rm E}^{(2)}$  & $R_{\rm E}^{(3)}$ 
  &   &    &  \\ 
\hline
B1152+199 & $\approx 0.8$ & $\cdots$  & $\cdots$   & $\cdots$   & $\approx 0$ 
      & G2 ignored   &  \\
B1359+154 & 0.36 & 0.72 & 0.29 & 0.29 & 0.50   &    &  \citet{Rus01} \\
B1608+656 & 0.53 & 1.06 & 0.29 & $\cdots$ & 0.64 &   &  \citet{Koo03} \\
B2045+265 & 1.06 & 2.12 & 0.09 & $\cdots$ & 0.11 &   &  \citet{McK07b} \\
B2114+022 & 0.89 & 1.78  & 0.46 & $\cdots$ &  0.42 & two-plane lensing & 
  \citet{Cha01} \\
J0134$-$0931 & 0.17 & 0.34 & 0.17 & $\cdots$ & 0.24 &   &\citet{KW03} \\
B1030+074 &  $\approx 0.8$ & $\cdots$ & $\cdots$ & $\cdots$  & $\approx 0$  
   &  G2 ignored  &   \\
B2108+213  & 1.74 & 3.48 & 0.51 & $\cdots$  & 0.55 & cluster/group ignored  &
   \citet{Mor08} \\
\hline
\end{tabular}

Notes: $R_{\rm E}^{(n)}$ -- $n$-th lensing galaxy's (G$n$'s) Einstein radius 
(ER);  $\Ri$ -- the average separation of the images from the primary galaxy 
(G1). All the values are given in arcseconds.
\end{table*}

Another important issue regarding the proper interpretation of the lens data 
is how to treat the multiple-lens systems. The operational definition for 
a multiple-lens system to be adopted in this work is a system in which the 
centres of distinct multiple lensing galaxies are found within 
the Einstein radius of the lens. For such a multiple-lens system 
care must be taken in interpreting the image separation. 
If secondary galaxies are much less massive than the primary galaxy and/or 
lie well outside the Einstein ring, 
then secondary galaxies may contribute little to the Einstein mass 
so that the image separation may well correspond to the
velocity dispersion of the primary galaxy up to small errors. 
Clearly, this appears to be the case at least for B1152+199 and B1030+074. 
On the other hand, if the lensing galaxies are of comparable mass 
and/or well within the Einstein radius as in interacting 
galaxies or a chance alignment (in projection) of galaxies at the same
redshift (as in a group) or at different redshifts, 
then the image separation cannot match well the velocity dispersion of the 
primary galaxy. This appears to be the case for the majority of the 
multiple-lens systems. Table~\ref{multi} summarises the Einstein radii of
the lensing galaxies in the multiple-lens systems.  
Indeed, the difference between the Einstein radius
of the total lens potential approximated by $\Ri$ and that of the primary
lensing galaxy $R_{\rm E}^{(1)}$ is quite large. This means that taking the
face values of $\twoRi$ of the multiple-lens systems would make a 
significant error in a statistical analysis of strong lensing. Simply ignoring
the image separations of the multiple-lens systems would not be a perfect
solution either because the primary galaxies of the multiple-lens systems might
be biased compared with `single' lensing galaxies. The average value of 
$R_{\rm E}^{(1)}$ for the 6 multiple-lens systems in Table~\ref{multi} is 
$0''.79$ while that for the rest of 24 systems is $\approx \twoRi/2 =0''.61$.
If B2108+213 were excluded assuming that the large separation is assisted by a
group/cluster, then the average value of $R_{\rm E}^{(1)}$ would be $0''.60$.
However, the measured velocity dispersion of G1 in B2108+213 appears to be 
consistent with the large separation without any significant contribution from
the group/cluster (\citealt{Mor08}) implying that B2108+213 may well be 
a valid data point.  Therefore, for the multiple-lens systems
we use $2R_{\rm E}^{(1)}$ instead of $\twoRi$ in our analyses. 
In doing so, we treat multiple-lens systems
as if they were single-lens systems but with corrected image separations.

\subsection{Model}

Suppose that there is a population of cosmological sources described by
a luminosity function or a number-flux density relation $N_{z_s}(>f_\nu)$
(number of sources with flux density greater than $f_\nu$ in a certain 
comoving volume). Then, for a source with observed redshift $z_s$ and flux 
density $f_\nu$ the probability that it is multiply-imaged 
(with a certain image multiplicity)
by intervening galaxies is, crudely speaking,
\begin{eqnarray} 
{\rm probability} & = & {\mbox{distance to the source}}  \nonumber \\
   &   & \times {\mbox{ number density of galaxies}}  \nonumber \\
   &   & \times {\mbox{ multiple-imaging cross section}} \nonumber \\ 
   &   & \times {\mbox{ magnification bias}}
\label{SLcrude}
\end{eqnarray}
 and can be formulated as follows in the order of the four
factors:
\begin{eqnarray}
p(z_s ,f_\nu ) & = & \int_0^{z_s} dz \left| \frac{c dt}{dz} \right| 
\int_{\Delta\theta_{\rm min}}^{\infty} d(\Delta\theta)  \nonumber \\
 &  & \times \int_0^{\infty} d\sigma \phi(\sigma,z) s(\sigma,z) 
\int_{\mu_{\rm min}}^{\infty} \frac{d\mu}{\mu}\mathcal{P}_\sigma(\Delta\theta,\mu)
  \nonumber \\ &  & \times 
  \left| \frac{dN_{z_s}(>f_\nu/\mu)}{df_\nu} \right| 
 \left|\frac{dN_{z_s}(>f_\nu)}{df_\nu} \right|^{-1},
\label{SLform}
\end{eqnarray}
where $t$ is cosmic time (i.e.\ proper time of a comoving galaxy), 
$\phi(\sigma,z)$ is the velocity dispersion function, 
$s(\sigma,z)$ is the cross section of a galaxy 
with velocity dispersion $\sigma$ at redshift $z$, and $\Delta\theta_{\rm min}$
and $\mu_{\rm min}$ are respectively the minimum allowed values of the 
separation and the magnification. In equation~(\ref{SLform}), 
$\mathcal{P}_\sigma(\Delta\theta,\mu)$ is the probability density distribution
of image separation ($\Delta\theta$) and magnification ($\mu$) within the 
multiple-imaging region for a galaxy of velocity dispersion $\sigma$. 
$\mathcal{P}_\sigma(\Delta\theta,\mu)$ depends on the lens model. For the 
simple case of the singular isothermal sphere (SIS) it is given by
\beq
\mathcal{P}_\sigma(\Delta\theta,\mu) = 
\delta(\Delta\theta-\Delta\theta_{\rm SIS}) \frac{2 \mu_{\rm min}^2}{\mu^3}
\label{psis}
\eeq
with 
\beq
\Delta\theta_{\rm SIS}=8\pi \frac{r_A(z,z_s)}{r_A(0,z_s)}
\left(\frac{\sigma}{c}\right)^2,
\label{thetsis}
\eeq
where $c$ is the speed of light in vacuum and  $r_A(z_1,z_2)$ is the 
angular-diameter distance between redshifts $z_1$
and $z_2$ in units of the Hubble radius.

We use the singular isothermal ellipsoid (SIE) lens model described by 
\citet{Cha03} to calculate lensing probabilities for a population of galaxies
given by  a VDF ({equation}~\ref{VDF}). We assume a mean projected mass 
density axis ratio of 0.7 ignoring the scatter.  The expressions for
absolute lensing probabilities, i.e.\ those including magnification biases, 
can be found in \citet{Cha03} while relative probabilities of image 
separations can be found in \citet{Cha05}. 
The total likelihood for lensing due to a certain population of galaxies is 
given by
\beq
\ln \mathcal{L}_{\rm tot} = 
      \ln \mathcal{L}_{\rm opt} + \ln \mathcal{L}_{\rm rad},
\label{Ltot}
\eeq
where $\mathcal{L}_{\rm opt}$ is the likelihood due to optically selected 
sources given by
\beq
\ln \mathcal{L}_{\rm opt} =
                 \sum_{k=1}^{N_{\rm U}^{\rm (opt)}} \ln [1 - p^{\rm (opt)}(k)]
            + \sum_{l=1}^{N_{\rm L}^{\rm (opt)}} \ln \delta p^{\rm (opt)}(l) ,
\label{Lopt}
\eeq
and  $\mathcal{L}_{\rm rad}$ is the likelihood due to radio-selected sources 
given by
\begin{eqnarray}
\ln \mathcal{L}_{\rm rad} & = &
 \left(  \sum_{k=1}^{N_{\rm U}^{\rm (rad)}} \ln [1 - p^{\rm (rad)}(k)] 
 + \sum_{l=1}^{N_{\rm L}^{\rm (rad)}} \ln \delta p^{\rm (rad)}(l) \right)
            \nonumber \\
     &   & + \left(   \sum_{j=1}^{N_{\rm IS}^{\rm (rad)}}
             \ln \delta p_{\rm IS}^{\rm (rad)}(j) \right).
\label{Lrad}
\end{eqnarray}
Then a ``$\chi^2$'' is defined by 
\beq
  \chi^2=-2 \ln \mathcal{L}_{\rm tot}.
\label{chisq}
\eeq
Here $p(k)$ is the integrated multiple-imaging probability,  
$\delta p(l)$ is the differential lensing probability of 
the specific separation, lens and source redshifts, 
and  $\delta p_{\rm IS}(j)$ is the relative image separation probability. 

For equation~(\ref{Lopt}) the Snapshot data are used. The lensed systems are
the 4 systems of the first part in table~\ref{lens} and 
$N_{\rm U}^{\rm (opt)}=502$. For equation~(\ref{Lrad}) the CLASS and PANELS data
are used. For the first parenthesis in equation~(\ref{Lrad}) the lensed systems
are the 13 systems of the 2nd part of Table~\ref{lens} and
 $N_{\rm U}^{\rm (rad)}=8945$. 
For the second parenthesis in equation~(\ref{Lrad}) the lensed systems are 
the 13 systems of the 3rd part of Table~\ref{lens}. 

We calculate lensing statistics for three kinds of galaxy population,
i.e. the early-type, the late-type and the total population.
For the population of a certain morphological-type we must include only 
the lensed systems of the same morphological-type lensing galaxies. 
However, for several lensed systems the lensing galaxy types have not yet been 
determined by observations nor could be judged otherwise.
These systems include Q1208+1011, B0128+437,  B0739+366 and B1555+375.
These systems all have relatively small image separations. Each of these 
systems may well be either early-type or late-type. 
Hence we include these systems for the 
statistics of both the early-type and the late-type populations but with
their contributions to the $\chi^2$ multiplied by a penalisation factor of 0.5. 

\section{Comparing the local VDFs with intermediate redshift ($0.3 \la z \la 1$) VDFs}

For the source redshift distributions of the CLASS and the
Snapshot the likely redshift of the lensing galaxy lies in $0.3 \la z \la 1$
(see Table~\ref{lens}). Because of this, if we constrain a non-evolving VDF 
using the lens data the constrained VDF would correspond to an averaged VDF over
the redshift range or a VDF at $z \sim 0.65$ assuming the
evolution is smooth in $z$. Comparing the constrained VDF of a certain 
galaxy population with the corresponding local VDF may then reveal the 
essential features of the evolution of the VDF.

We constrain the VDF of the form given by equation~(\ref{VDF}) for both the
type-specific VDFs and the total VDF. The local total VDF is well-fitted by
the six-parameter function of equation~(\ref{VDFtot}) rather than the function
 of equation~(\ref{VDF}) (see \S 2.4). However, the number statistics of the
strong lens sample is not strong enough to warrant the six-parameter function.
When all four parameters of the VDF (equation~\ref{VDF}) are allowed to vary,
 the confidence regions in the parameter space are broad because of parameter 
degeneracies. However, the VDF itself is constrained well within plausible
ranges of the parameters. Hence we may fix the low-velocity end slope $\alpha$ 
to the local value to break the parameter degeneracy without 
significantly altering the possible range of the VDF.

\begin{figure*}
\begin{center}
\setlength{\unitlength}{1cm}
\begin{picture}(14,10)(0,0)
\put(-2.5,12.){\includegraphics{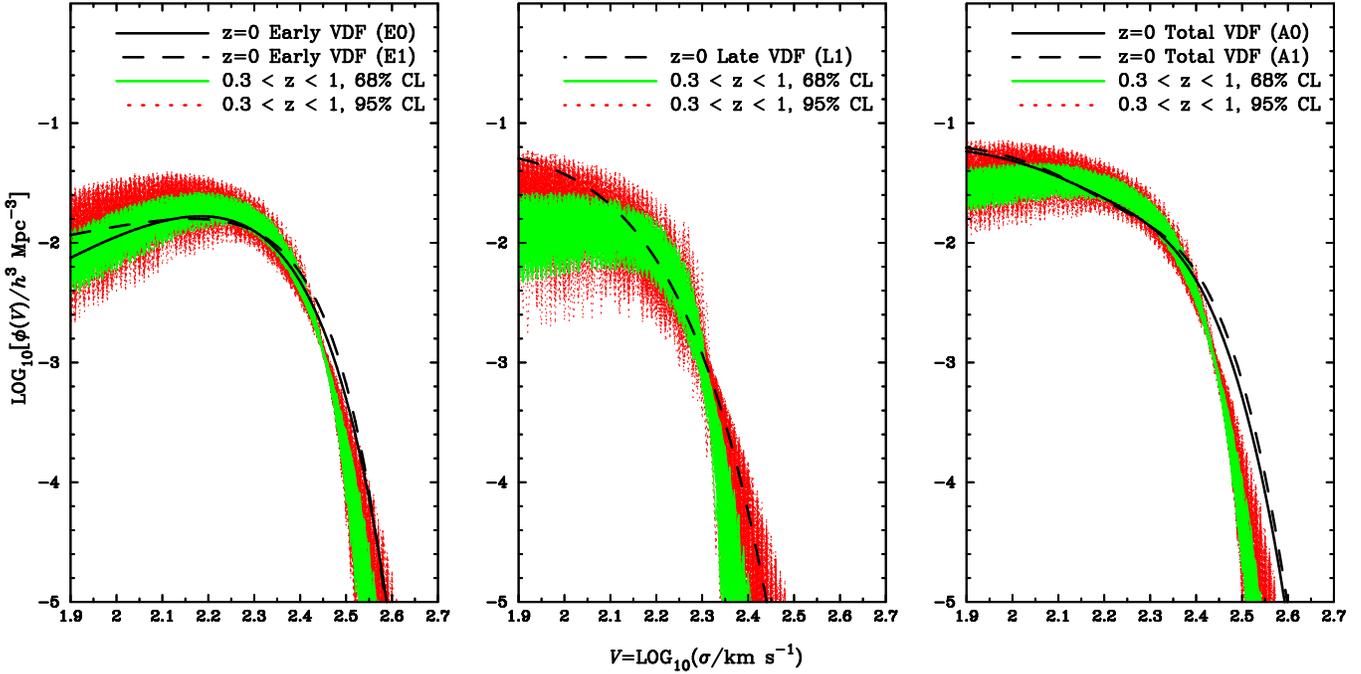}}
\end{picture}
\caption{
The strong lensing constrained intermediate-redshift ($0.3 \la z \la 1$) VDFs
are compared with the local VDFs (Fig.~\ref{VDFs} and table~\ref{GFs}). 
These simple comparisons reveal the trends of the evolutions without detailed
modelling. The green and red regions are respectively the 68\% and the 95\%
confidence limits.
}
\label{VDFcomp}
\end{center}
\end{figure*}

Fig.~\ref{VDFcomp} shows the 68\% and 95\% confidence limits (CLs) of the
constrained VDF for the total and the type-specific galaxy 
populations. In Fig.~\ref{VDFcomp} the local VDFs are overplotted and compared
with the constrained VDFs. We notice the following trends for the evolutions of
the VDFs. First, the number density of highest velocity dispersion galaxies
appears to evolve most as can be seen from both the total VDF and the early-type
VDF. Second, the lower velocity dispersion part of the early-type VDF does not
show any signature of evolution. Finally, the late-type VDF evolves little 
overall.

\section{Constraints on the evolutions of the velocity dispersion functions: parametric modelling}

We study the evolution of the VDF through a parametric approach.
We parametrise the evolution of the number density and the velocity 
dispersion as
\beq
 \phi^*(z)  =   \phi^*_0 10^{Pz}; \hspace{1em}
 \sigma^*(z)  =   \sigma^*_0 10^{(Q/4)z}, 
\label{NVevol}
\eeq
and the evolution of the shape of the VDF as
\beq
\alpha(z)  =  \alpha_0 \left(1 + k_{\alpha} \frac{z}{1+z}\right); \hspace{1em}
 \beta(z)  =  \beta_0 \left(1 + k_{\beta} \frac{z}{1+z} \right).
\label{ABevol}
\eeq
Here $z$ is the cosmological redshift and $z/(1+z)=1-(a/a_0)$ (where 
$a$ and $a_0$ are the cosmological scale factors at redshift $z$ and zero
respectively). Parameters $P$ and $Q$ in equation~(\ref{NVevol}) are 
parametrised such that they have the same sensitivity to the strong lensing 
optical depth (the absolute lensing probability).  
The positivity of $\alpha$ and $\beta$ in 
equation~(\ref{ABevol}) requires that $k_{\alpha}>-1$ and $k_{\beta}>-1$.
Notice that parameters $Q$, $k_{\alpha}$ and $k_{\beta}$ have 
sensitivities to the image separation distributions 
as well as the lensing optical depth while parameter $P$ has sensitivity
only to the lensing optical depth.

The present lens sample is limited to
the image separation $\Delta \theta \ga 0''.3$. Consequently, the data
cannot constrain the shape of the VDF for 
$\sigma \la 95\kms$.\footnote{For a lens redshift of 0.6 and a source redshift
of 2. $\Delta \theta = 0''.3$ corresponds to $\sigma \approx 95\kms$ for the
assumed cosmology.} Hence the constraints from this study are valid
only for $\sigma \ga 95\kms$ and outside this range the results must be
regarded as extrapolations.
Because of this limitation of the lens data and the small number statistics
parameter $k_\alpha$ is ill-constrained by the data. Moreover, 
for a given shape of the VDF there are parameter degeneracies as pointed out
in \S4. Hence it is necessary to impose a constraint on the parameters.
We may consider $k_\alpha=0$ or $k_\alpha=k_\beta$ as a default choice. 
The former choice 
corresponds to a constant shape for the low velocity end slope while 
the latter corresponds to a constant faint-end slope since 
$\alpha_{\rm L}=\alpha(z)/\beta(z)-1$ (equation~\ref{corr}).
In practise these two choices make little difference for the present data.
We adopt the latter constraint because the intermediate-redshift VDFs 
constrained in \S 4 appear to be steeper both at the low velocity and the
high velocity ends.

\begin{table}
\caption{Evolution parameters for the velocity dispersion 
functions \label{evparam}}
\begin{tabular}{cccc}
\hline
VDF & $P$ & $Q$ & $k_\beta$   \\
\hline
\multicolumn{4}{c}{Early-type} \\
\hline
  E0 & $+0.02_{-0.32}^{+0.38}$ & $-0.08_{-0.48}^{+0.33}$
         & $+0.12_{-0.57}^{+0.67}$ \\
  E1 & $-0.15_{-0.34}^{+0.46}$ & $-0.01_{-0.38}^{+0.21}$ 
         & $+0.65_{-0.97}^{+1.05}$ \\
\hline
\multicolumn{4}{c}{Late-type} \\
\hline
 L1 & $-0.56_{-0.66}^{+0.74}$ & $+0.40_{-0.84}^{+0.56}$ 
         & $+0.34_{-0.81}^{+0.93}$ \\
\hline
\multicolumn{4}{c}{All} \\
\hline
 A0 & $+0.73_{-0.28}^{+0.23}$ & $-1.87_{-0.63}^{+0.86}$
         & $-0.71_{-0.16}^{+0.30}$ \\
 A1 & $+0.81_{-0.23}^{+0.21}$ & $-2.18_{-0.56}^{+0.79}$ 
         & $-0.77_{-0.13}^{+0.25}$ \\
\hline
\end{tabular}

\end{table}

\subsection{Early-type galaxies}

\begin{figure}
\begin{center}
\setlength{\unitlength}{1cm}
\begin{picture}(9,9)(0,0)
\put(-0.7,-2.){\includegraphics{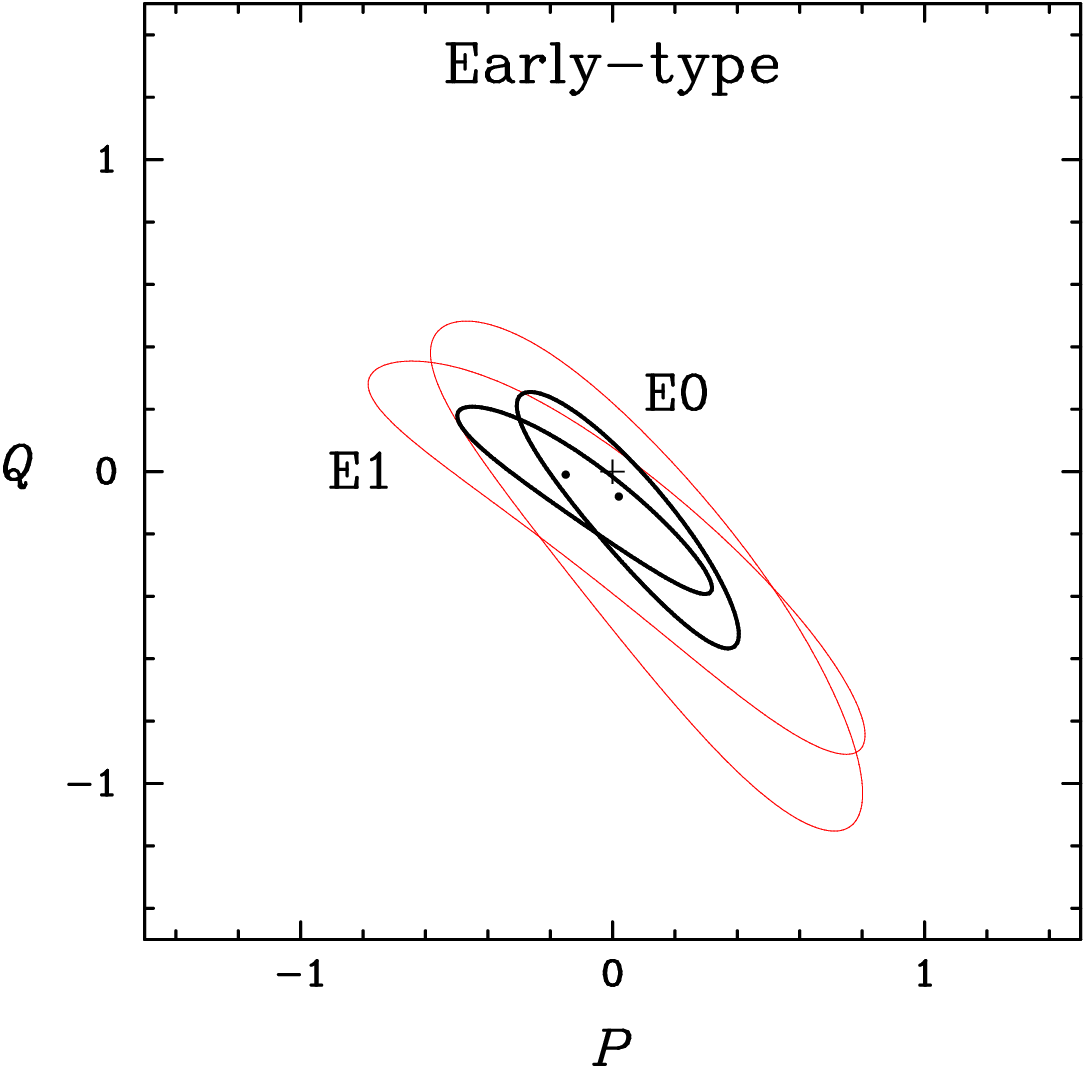}}
\end{picture}
\caption{
The 68\% and the 95\% confidence limits in the evolutionary parameters of
the early-type velocity dispersion function. 
The integrated number density evolves as $10^{Pz}$ 
while the velocity dispersion parameter $\sigma^* \propto 10^{Qz/4}$.
The parameter for the evolution of the shape is not shown here but can be
found in table~\ref{evparam}. 
}
\label{Eplane}
\end{center}
\end{figure}

{Fig.}~\ref{Eplane} shows the 68\% and 95\% CLs in the $P$-$Q$ plane for two 
local early-type VDFs. The fitted parameter values can be found in 
Table~\ref{evparam}. The early-type VDFs (both the \citet{Cho07} measured 
E0 VDF and the Monte Carlo generated E1 VDF) evolve little overall.
The overall little evolution of the early-type VDF
is consistent with previous constraints from strong 
lensing  statistics (e.g. \citealt{CM03,Ofe03}).

\begin{figure}
\begin{center}
\setlength{\unitlength}{1cm}
\begin{picture}(8,11)(0,0)
\put(-0.7,-0.8){\includegraphics{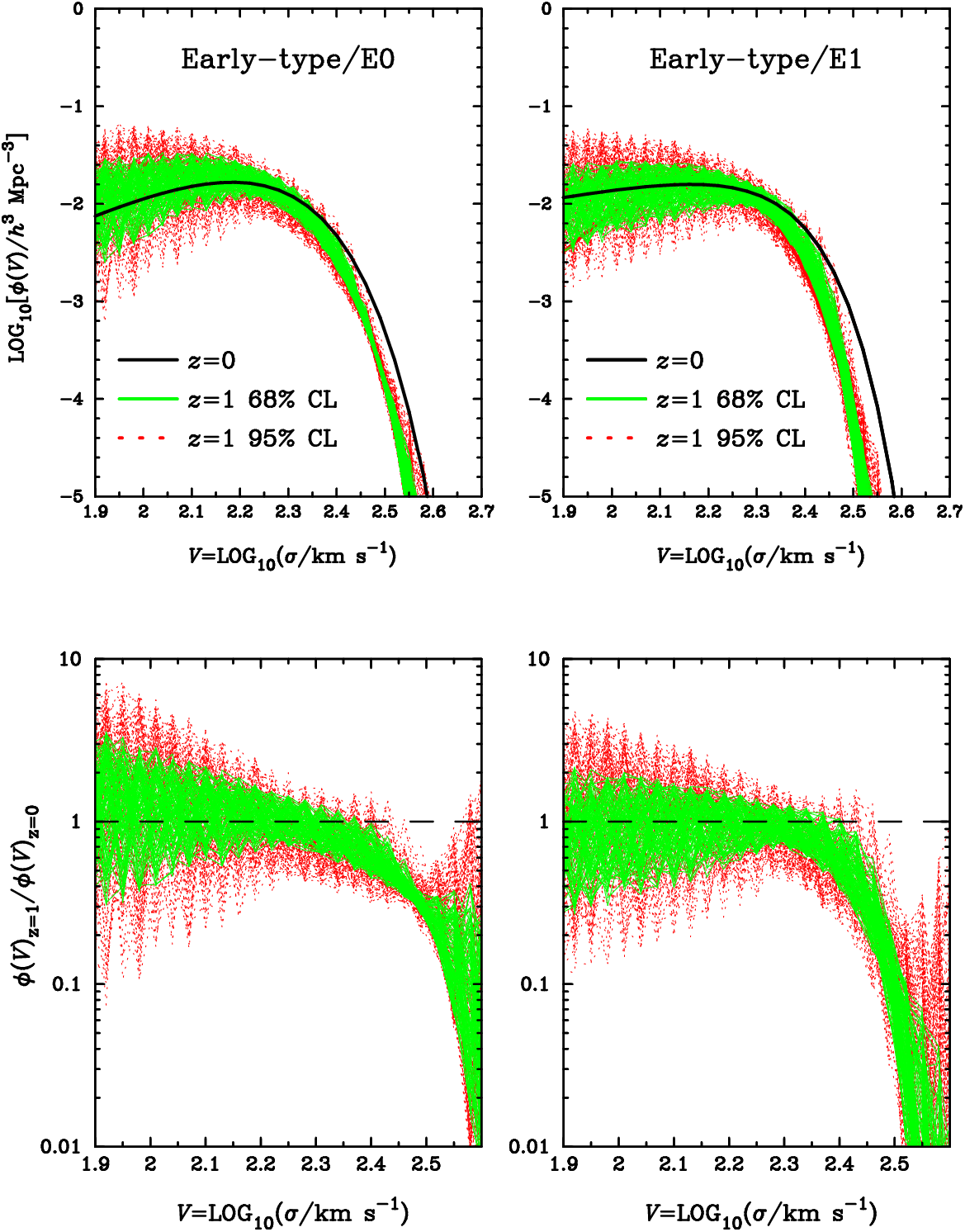}}
\end{picture}
\caption{
The early-type VDFs at $z=1$ are compared with the local counterparts. 
The $z=1$ VDFs are the functions projected using the constrained evolution
 parameters (table~\ref{evparam}).
The green and red regions show the 68\% and 95\% confidence limits due to
the uncertainties in the constrained parameters.
These figures clearly show the differential number density evolution for
early-type galaxies of high velocity dispersion in which the evolution is
greater at a higher velocity dispersion.
}
\label{Eevol}
\end{center}
\end{figure}

However, more detailed nature of the evolution as a function of velocity 
dispersion can be found in {Fig.}~\ref{Eevol} in which the local VDFs are 
compared with the corresponding projected VDFs at $z=1$. 
The main features of {Fig.}~\ref{Eevol} may be summarised as follows. 
First, the number densities of early-type galaxies of 
large velocity dispersions ($\sigma \ga 230-250 \kms$) have not only changed 
significantly from $z=1$ to $z=0$ but by an increasingly larger factor for a 
larger velocity dispersion.  The dependence of the number density evolution
on velocity dispersion appears to be real because a constant shape is clearly
disfavoured by data. In particular, the number density of the largest velocity
dispersion galaxies ($\sigma \ga 300 \kms$)  has increased by a large factor
($\ga 3$) since $z=1$. 
 Second, for early-type galaxies of typical velocity dispersions and lower
($\sigma \la 230-250 \kms$) there is no statistically significant change in 
number density from $z=1$ to $z=0$. 
This explains why lens statistics gives the ``right'' cosmology
assuming no evolution in the number density and the shape (see \S 6.2).

Finally, we must bear in mind that the above results for the early-type
population suffer from systematic errors arising from the lack of the
 knowledge of the lensing galaxy morphological type for several lens systems
(see \S 3). The results for the late-type population shown below 
are prone to the same problem. However, the results for the evolution of
the total VDF (\S 4.3) are free from the problem.

\subsection{Late-type galaxies}

\begin{figure}
\begin{center}
\setlength{\unitlength}{1cm}
\begin{picture}(8,8)(0,0)
\put(-0.7,-2.3){\includegraphics{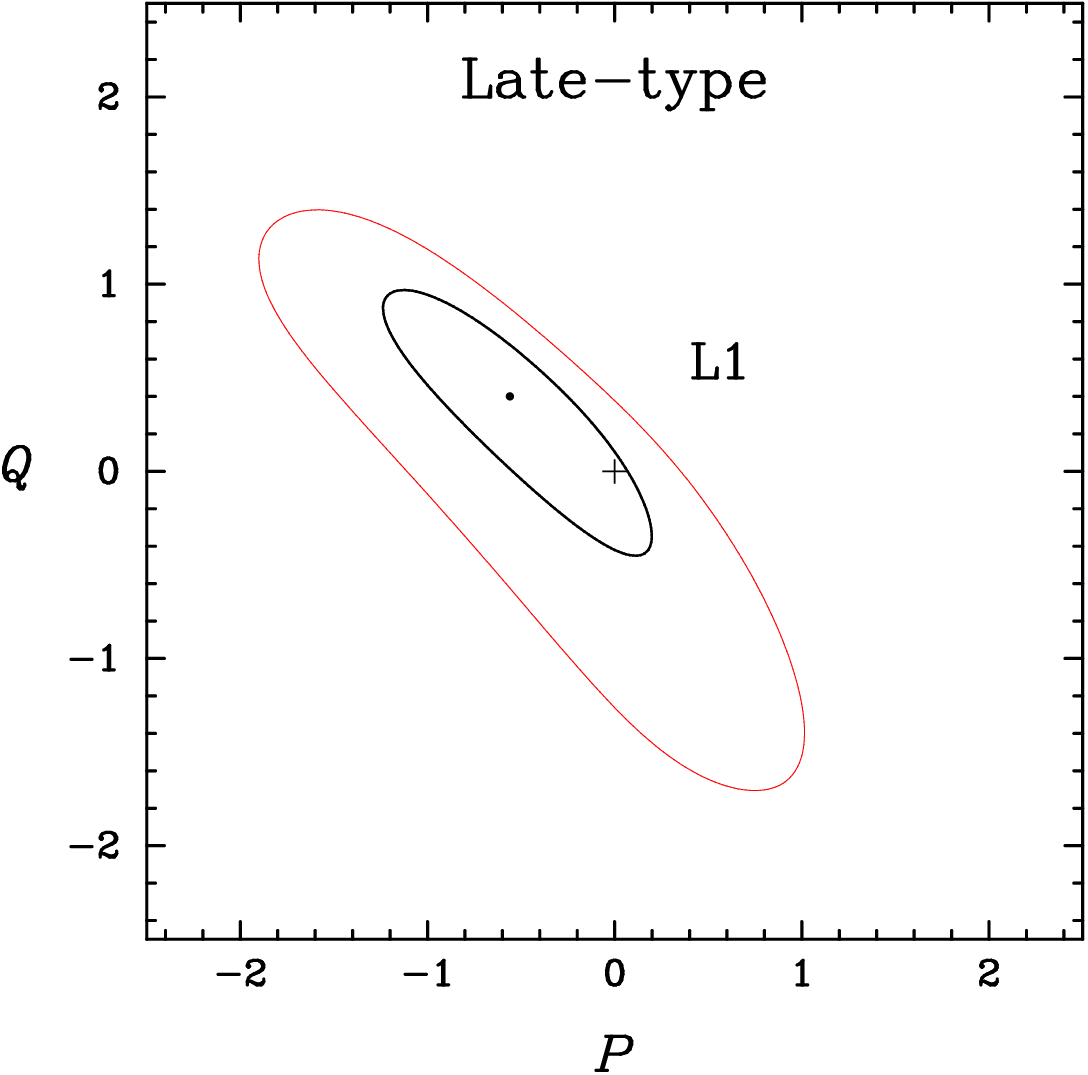}}
\end{picture}
\caption{
The same as Fig.~\ref{Eplane} but for the late-type population of galaxies.
}
\label{Lplane}
\end{center}
\end{figure}

\begin{figure}
\begin{center}
\setlength{\unitlength}{1cm}
\begin{picture}(8,11)(0,0)
\put(-0.7,-0.6){\includegraphics{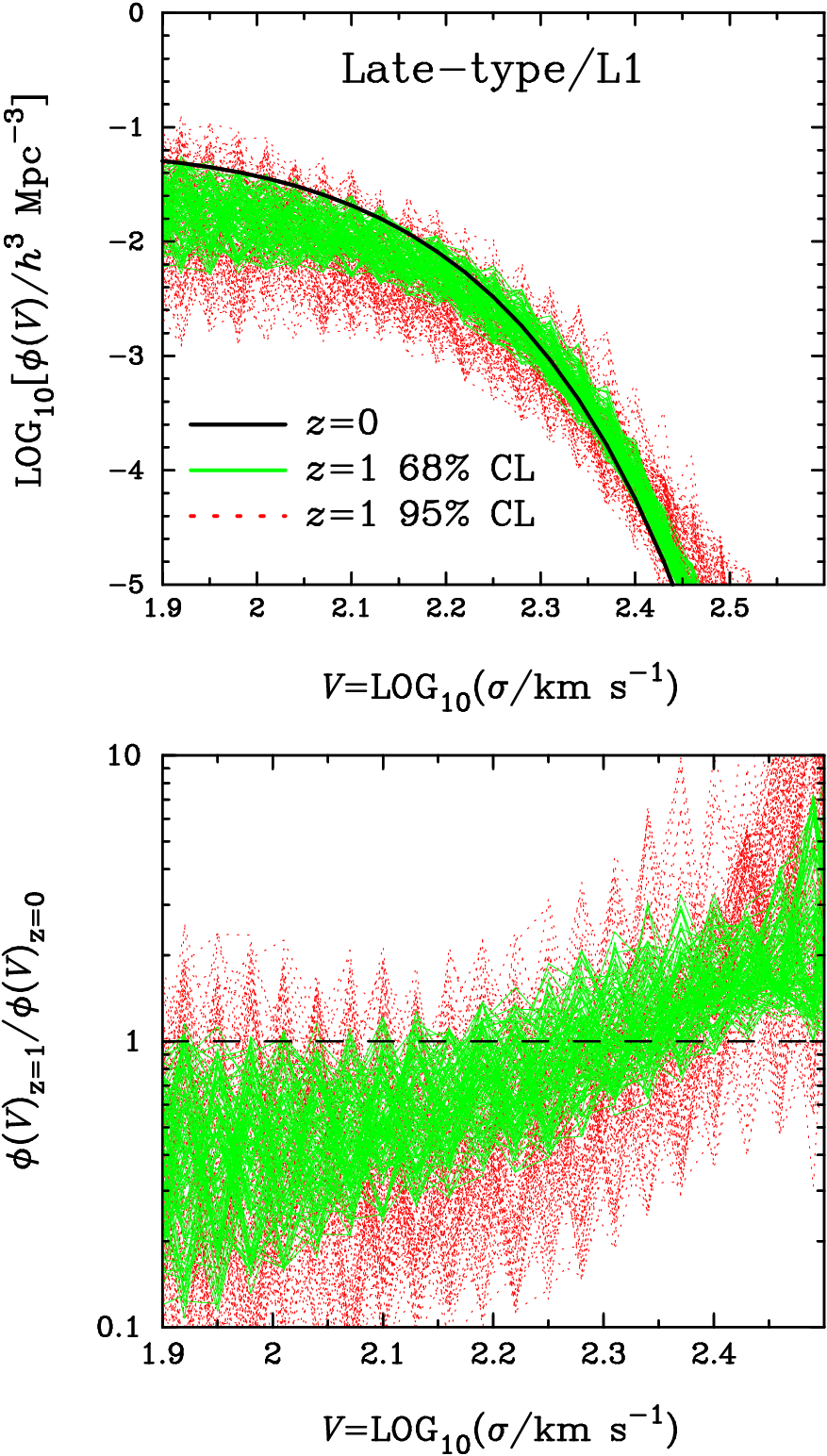}}
\end{picture}
\caption{
The same as Fig.~\ref{Eevol} but for the late-type population of galaxies.
}
\label{Levol}
\end{center}
\end{figure}

The evolutionary behaviours of the late-type VDF can be found in 
Figs.~\ref{Lplane} \& \ref{Levol}.
Overall, the late-type VDF is consistent with no evolution at the 68\% 
confidence level.
Furthermore, in contrast to the early-type VDF there is no statistically 
significant differential evolution for the late-type VDF. 
In particular, the number density of the most massive late-type galaxies 
does not change significantly between $z=1$ and $z=0$.

\subsection{All galaxies}

\begin{figure}
\begin{center}
\setlength{\unitlength}{1cm}
\begin{picture}(9,11)(0,0)
\put(-0.5,-1.){\includegraphics{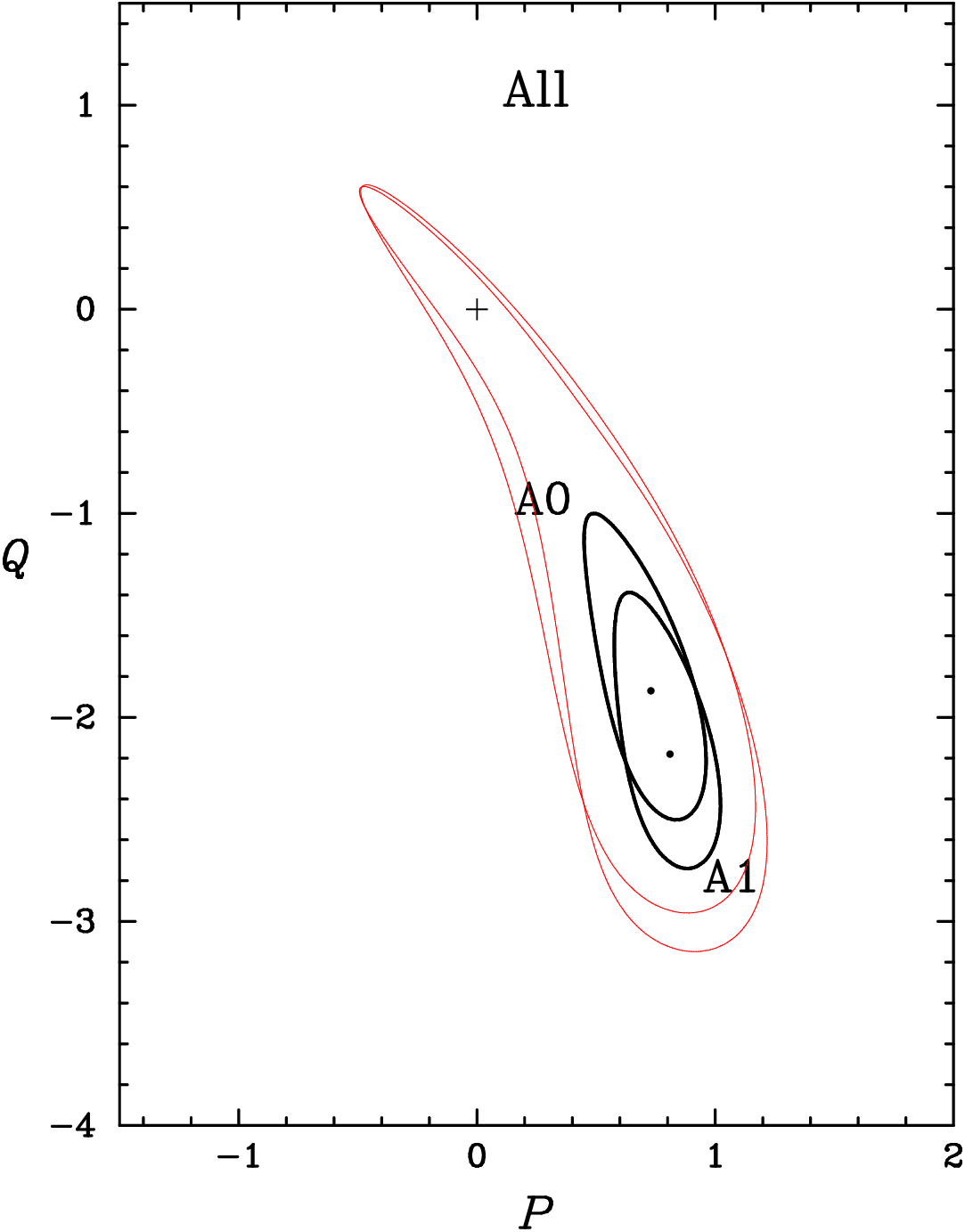}}
\end{picture}
\caption{
The same as Fig.~\ref{Eplane} but for the total population of galaxies.
}
\label{Aplane}
\end{center}
\end{figure}

We study the evolution of all galaxies as a whole using the
total VDFs  A0 and A1 in Table~\ref{GFs}. One advantage of constraining
the evolution of the total VDF is that it is not affected by the uncertainties 
in the identification of the lensing galaxy type for several systems. 
However, parametric modelling of the total VDF is less straightforward
because the total VDF has more parameters (equation~\ref{VDFtot}) than
the type-specific VDF (equation~\ref{VDF}). Namely, there arise more severe 
degeneracies amongst the evolutions of the parameters. To break the parameter
degeneracies we choose to fix the parameters of the correction term in 
equation~(\ref{VDFtot}), i.e. $\alpha'$ and $\varepsilon$.
{Fig.}~\ref{Aplane} shows the 68\% and 95\% CLs in the $P$-$Q$ plane. 
The fitted parameter values can be found in Table~\ref{evparam}.
Notice that the fitted values of the evolution parameters are quite different 
from those of the early-type VDF because of the differences in the VDF 
functions as well as the data used.

\begin{figure}
\begin{center}
\setlength{\unitlength}{1cm}
\begin{picture}(8,11)(0,0)
\put(-0.7,-0.8){\includegraphics{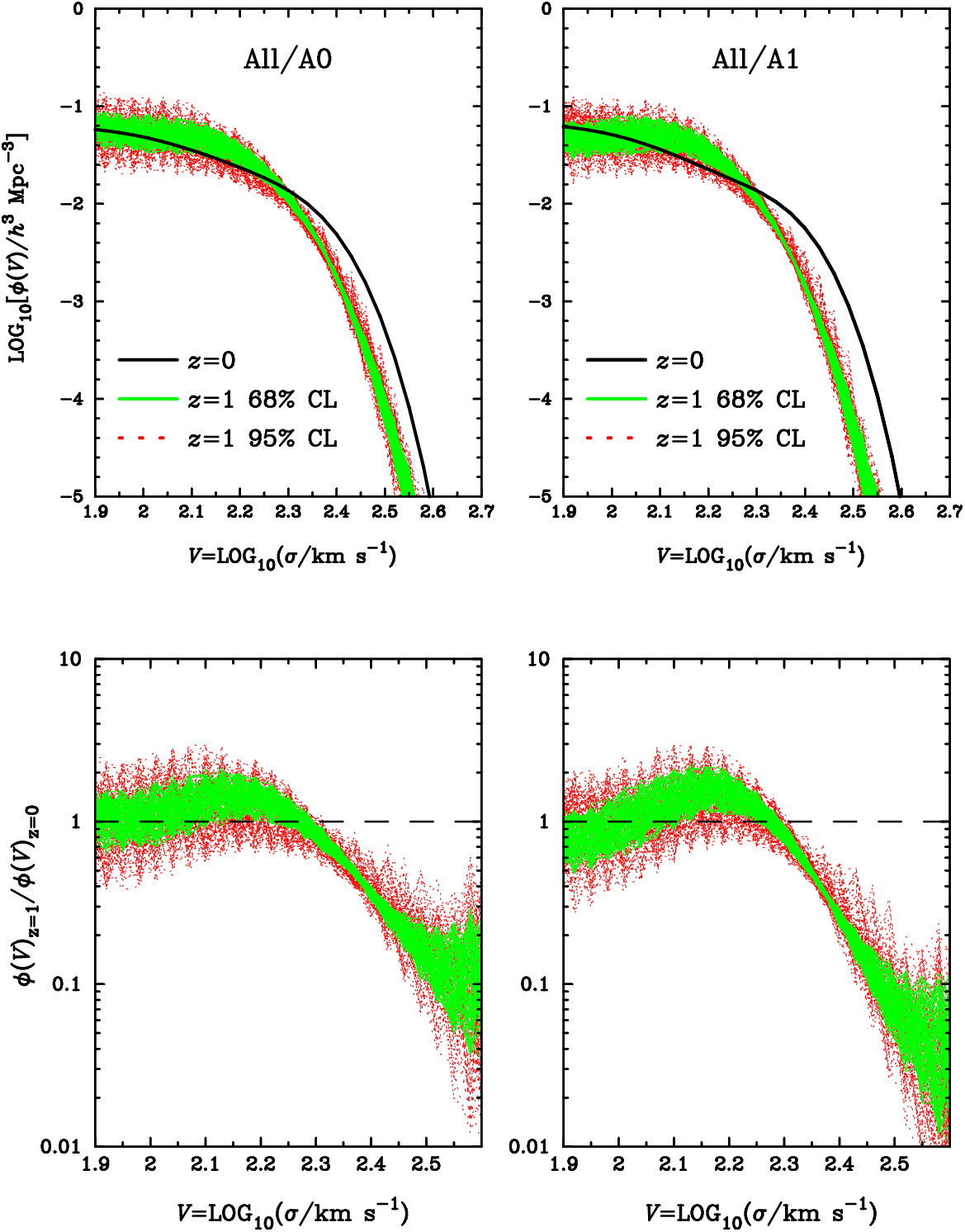}}
\end{picture}
\caption{
The same as Fig.~\ref{Eevol} but for the total population of galaxies.
}
\label{Aevol}
\end{center}
\end{figure}

In {Fig.}~\ref{Aevol} the total VDFs at $z=0$ are compared with the
projected total VDFs at $z=1$. The inferred evolutionary trends of the
total VDFs are consistent with those of the type-specific VDFs. There are 
only relatively minor quantitative differences between the total VDF and the 
type-specific VDFs in the detailed nature of the evolution. 
Most noticeably, the number density evolution starts to become important at 
$\sigma \approx 200 \kms$ from the inferred evolution of the total VDF
(whereas it starts at $\sigma \approx 250 \kms$ for the early-type VDF) and
the degree of evolution for large velocity dispersion galaxies is greater
for the total VDF than the early-type VDF. These minor quantitative
differences are due to the differences both in the functional form of the
VDF and the data used between the total VDF and the early-type VDF.
The differences may reflect a possible range of errors.  
Despite the minor quantitative differences in the evolution between the total 
VDF and the early-type VDF, the overall trends of the evolution for 
large velocity dispersion galaxies are essentially the same.

\section{Discussion}

\subsection{Possible sources of systematic errors}

\subsubsection{(The interpretations of) the strong lensing data}

The results presented in \S5 are based on the most likely interpretation of the
strong lensing data as judged by the author. However, the present data do not
exclude alternative interpretations of the strong lensing systems. We assess
here the greatest possible effects on galaxy evolution of alternative 
interpretations of the strong lensing data.

\begin{figure}
\begin{center}
\setlength{\unitlength}{1cm}
\begin{picture}(9,14)(0,0)
\put(-1.9,-0.5){\includegraphics{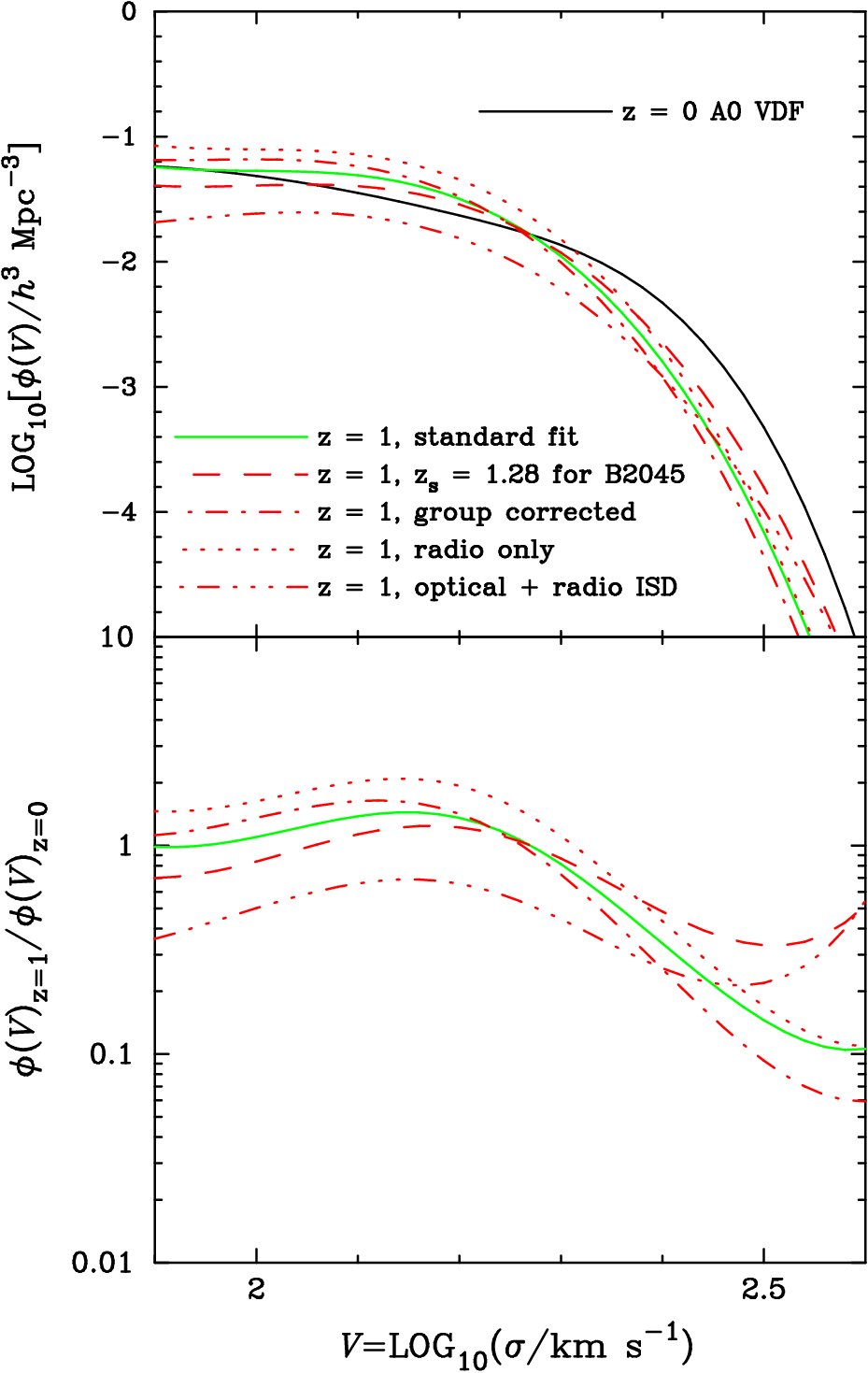}}
\end{picture}
\caption{
Possible systematic errors in the evolution of the VDF.
The green solid line is the best-fit for the A0 VDF as shown in 
Fig.~\ref{Aevol}. The red dashed line is the fit assuming that the source 
redshift $z_s =1.28$ is adopted {\it and} interpreted to give 
$\sigma \approx 384 \kms$ for the lensing galaxy in B2045+265.
The red dash-dot line is the fit assuming that the surrounding groups
contributed significantly to the image separations and the cross sections for
several systems (see the texts in \S 6.1.1). The red dotted 
and dash-dot-dot-dot lines are the fits respectively including only 
the CLASS sample and only the Snapshot sample for the absolute lensing 
probability.
}
\label{VDFalt}
\end{center}
\end{figure}

First, it is possible that the source redshift for B2045+265 is $z_s=1.28$ and
accordingly the velocity dispersion of the lensing galaxy is 
$\sigma \approx 384 \kms$. We judge that this interpretation is unlikely 
based on the following arguments. First of all, the measured velocity 
dispersion of the lensing galaxy as reported by \citet{Ham05} is 
$\sigma = 213\pm 23 \kms$ which is much lower than the implied SIS velocity
dispersion. If the measurement is correct and the source redshift is $z_s=1.28$,
this system would be an extremely rare case in which the SIS velocity dispersion
is much larger than the measured stellar velocity dispersion 
($\sigma_{\rm stars}/\sigma_{\rm SIS}=0.55\pm 0.06$).
For the Sloan Lens ACS Survey (SLACS) sample of 53 early-type lens galaxies 
with $170 \kms < \sigma_{\rm stars} < 400 \kms$ \citet{Bol08} find 
$0.7 < \sigma_{\rm stars}/\sigma_{\rm SIS} < 1.3$ with a mean value close to 
unity. Secondly, according to \citet{Ham05} the lensing galaxy in B2045+265
lies in the fundamental plane with the measured velocity dispersion implying
that it is a normal elliptical galaxy with the `correct' velocity dispersion.
Thirdly, the source redshift $z_s=1.28$ for B2045+265 would give
a much worse $\chi^2$ for the model fit ($\Delta\chi^2 > 12$ for the evolution
model of the total VDF). This means that the source redshift $z_s=1.28$ for 
B2045+265 is not in harmony with all the other lens systems. 
Finally, the suggestion of  $z_s=1.28$ is based only on a single broad 
emission line (\citealt{Fas99}). This is far from
a secure redshift from the observational point of view alone.
Nevertheless, if we consider $z_s=1.28$ and so take $\sigma = 384\kms$ ignoring
the \citet{Ham05} measurement of $\sigma = 213\pm 23 \kms$  for B2045+265,
the evolutions of the early-type VDF and the total VDF would change
significantly at the high end of velocity dispersion. In Fig.~\ref{VDFalt}
the total VDF A0 at $z=0$ is compared with its projected VDF at $z=1$ with 
$z_s=1.28$ for B2045+265. The ratio $\phi_{z=1}/\phi_{z=0}$ decreases up to
$\sigma \sim 300\kms$ but then turns upward so that it approaches near unity
within the 68\% CL (although it is not shown)  as
$\sigma$ reaches $400 \kms$. Namely, in this model the most massive
early-type galaxies are already in place at $z=1$ and there is little change
in their number density since then. Overall the number density of early-type
galaxies as a function of velocity dispersion is consistent with no evolution 
at 95\% confidence between $z=1$ and $z=0$.
This is in stark contrast to the results presented in \S 5
without $z_s=1.28$ for B2045+265 that require the greater evolution at
the larger velocity dispersion for $\sigma \ga 200 \kms$. 
Given this potentially large systematic error, 
it is important to measure more reliably the source redshift and 
confirm the stellar velocity dispersion of \citet{Ham05} for  B2045+265.

Second, the environments of the lensing galaxies may cause biases in 
inferring the velocity dispersions of the lensing galaxies from the observed 
image separations. We corrected the effects of the secondary (and tertiary)
lensing galaxies within the image regions based on the results of detailed 
lens modelling found in the literature (Table~\ref{multi}). 
However, we ignored the possible effects of groups or clusters surrounding 
(and galaxies nearby) the image regions. It is thus possible that 
the velocity dispersions of the lensing galaxies were overestimated for 
some fraction of the lenses. 
There have been no systematic observations on the environments of the CLASS or
the Snapshot lenses. However, a recent study of the environments of the SLACS
lenses finds that typical contributions from the environments of the `normal'
lenses (i.e.\ those without very close companions) are less than a few per cent
in the lensing mass density (\citealt{Tre08}). For other several lens systems
(including some CLASS and Snapshot lenses)
with identified surrounding groups/clusters recent studies find that the 
environments contribute typically less than 10\% to the convergence of 
the lens potential (\citealt{Mom06,Aug07,Fas08}). 
Since the lens surface density is proportional to the velocity dispersion 
squared, neglection of the environmental contributions would cause errors 
in the velocity dispersion at most about 3\%. Theoretical simulations by 
\citet{KZ04} predict somewhat larger error of 6\% for the velocity dispersion.
To assess a maximum possible systematic error arsing
from neglecting the lens environments we consider lowering the observed image
separation by 25\% for the nine lens systems with a group (or groups) along the 
sight line (see table~\ref{lens}) assuming that the implied velocity dispersion
 is biased by 5\%. The evolution of the total VDF A0 for this case can be 
found in Fig.~\ref{VDFalt}. The best-fit for this case gives a better
$\chi^2$ ($\Delta\chi^2 = -2.5$) compared with the case that the surrounding 
groups contribute nothing to the image separations. 
 However, the evolutionary behaviour of the VDF is little changed.

\subsubsection{Statistical properties of sources}

For the statistical analysis of strong lensing the properties of
the source population are of crucial importance. The statistical properties
of the source population influence the lensing probabilities through its
redshift distribution and number-magnitude (or number-flux density) relation
or through an evolving luminosity function collectively. Consequently, any
error in the source properties can bias the results based on strong lensing
statistics. For the present work, we use two distinct populations of sources,
namely optical quasars and flat-spectrum radio sources (CLASS sources). 
For the optical quasar population the source redshifts are all known 
and the number-magnitude relation is also known empirically. For the
flat-spectrum radio source population the source properties are less certain.
For most of the radio sources the redshifts are not known and the inferred
redshift distribution for the source population was largely based on 27
measured redshifts for a subsample of sources with 5~GHz flux densities
between 25~mJy and 50~mJy (\citealt{Mar00}). Since lensed sources can
come from flux densities below 25~mJy, it has been assumed for the previous
analyses of the CLASS lensing that the redshift distribution is unchanged 
for sources below 25~mJy. Hence it is possible that the present work suffers 
from a systematic error in the CLASS parent source redshift distribution.
 However, \citet{McK07a} reports (a preliminary result) that the mean redshift 
for a sample of flat-spectrum sources with 5~GHz flux densities between 
 5~mJy and 15~mJy is consistent with that for sources of higher flux densities
by \citet{Mar00}. The number-flux density relation of the flat-spectrum
radio sources is well measured above 30~mJy at 5~GHz because it is based on
a large number ($\sim 10^4$) of source counts. It is less certain
below 30~mJy because such a large number of sources are not available. 
Nevertheless, \citet{McK07a} gives a measurement of the number-flux density 
relation below 30~mJy based on a sample of 117 sources down to 
5~mJy and confirms a previous estimate reported by \citet{Cha02}. 

As discussed above there is no observational indication that the source 
properties adopted in this work are grossly biased. However, the uncertainties 
in the CLASS source properties clearly require further observational studies.
It is thus of interest to see whether the two independent statistical lens 
samples, namely the Snapshot sample and the CLASS statistical sample, give
agreeing results. The two independent results can provide a cross check of each 
other. Unfortunately, the Snapshot sample contains only four lenses so that
it has a weak statistical power in constraining the evolution of the VDF.
In particular, the Snapshot data alone cannot constrain the shape of the VDF
at all. We thus include the image separation distribution (ISD) of all 26 radio
lenses for the Snapshot fit. Notice that the ISD of the radio lenses does not
depend on the statistical properties of the CLASS parent source population
so that the inclusion of the radio ISD does not bias the result of the Snapshot
fit. In Fig.~\ref{VDFalt} the radio and the Snapshot best-fit results are
compared. Although not shown in the figure the 68\% CLs of the 
two results overlap so that they are consistent with each other. However,
the best-fit results indicate that the Snapshot sample requires more  
evolution overall but less evolution at the high mass end than the CLASS sample.

\subsubsection{Galaxy mass profiles}

The results of this work are based on the assumption that the galactic mass 
distribution follows the isothermal distribution in the inner region of the
galaxy probed by strong lensing (within about the effective radius).
This assumption naturally leads to the equality of the observed stellar
velocity dispersion and the SIE model velocity dispersion. Conversely,
if the SIE assumption is systematically violated the lens statistics
cannot reliably constrain the observed local stellar velocity dispersion.
Of course, not all galaxies will closely follow the isothermal mass profile.
However, the deviations of individual galaxies from the isothermality would not
matter much statistically as long as the average profile is close to the 
isothermal.

Systematic errors from the isothermality assumption can arise in the 
following two contexts.
First, if the mass profile varies systematically depending on the velocity
dispersion, e.g. a higher velocity dispersion galaxy follows a shallower 
profile on average (or vice versa), then the results of this work can be
systematically biased. The solution to this potential problem appears to be 
provided in part by the SLACS study of early-type galaxies through
a combination of lens modelling and stellar velocity dispersion measurement 
of lensing galaxies. \citet{Bol08} finds through an analysis of 53 
massive early-type galaxies that the mass profiles of early-type
galaxies scatter around the isothermal profile but the average profile
is close to the isothermal profile independent of the velocity dispersion
between $175\kms$ and $400\kms$. The \citet{Bol08} finding applies to the
upper velocity dispersion part to the peak velocity dispersion ($\sim 160\kms$)
of the early-type VDF. For the early-type galaxies of the peak velocity 
dispersion and lower, it remains unanswered  whether the isothermality 
assumption is systematically unbiased. However, down to the velocity dispersion
limit ($\sigma \approx 95\kms$) sensitive to the strong lensing data
(remember that this study is limited by the image separation range)  
most early-type galaxies are likely to be normal elliptical/lenticular 
galaxies with little contamination of dwarf ellipticals/lenticulars
(see, e.g., \citealt{deR05}).
Hence it is likely that the isothermality assumption remain not so bad for
low velocity dispersion galaxies within the range probed by this work.
For the late-type galaxies it is expected  that the mass profile becomes 
systematically shallower in the inner region 
($\la R_{\rm d}$\footnote{The disk scale length in the exponential disk.}) 
as the luminosity decreases for sub-$L_*$ galaxies according to the rotation 
curves of spiral galaxies (\citealt{Per96}).  However, for the range of
galaxies probed by the strong lensing data (circular rotation speed 
$V_c \ga 134\kms$) the deviations are not likely to be too large.
The effects of the neglected errors can be understood as follows. 
The assumed constant rotation curve for a galaxy whose rotation speed actually
declines toward the galactic centre will inflate the lensing mass and so the 
lensing cross section. This inflated lensing cross section will then lead an
underestimate of the number density so as to keep the observed lensing 
occurrences. Consequently, the evolution of the number density for the 
late-type population can be overestimated. We postpone the quantification
of this systematic error to a future work although it may not be warranted
because of the small number statistics of late-type lenses in the present data.
 
Second, if the average mass profile of galaxies deviates systematically from
the isothermal as $z$ increases, then the results on the galaxy number density 
evolution from this work can be biased. 
The total (luminous plus dark) mass profiles of galaxies at intermediate and 
higher redshifts can be best constrained by modelling strong lens systems 
provided with reliable and stringent observational constraints.
The best studied examples in strong lensing appear to suggest that the average
total mass profile of galaxies is isothermal up to 
$z \sim 1$.\footnote{It is now well established that lensing galaxies at 
intermediate redshifts $z \sim 0.3-0.7$ have isothermal (or close to isothermal)
mass profiles. The question here is whether the isothermality would hold up to 
$z=1$.}
\citet{Koo06} carry out a systematic modelling of SLACS lenses and find that 
the mass profiles of gravitational lens galaxies are 
close to isothermal regardless of the lens redshifts between $z=1$ and $z=0$.
The \citet{Koo06} sample contains 6 lenses for $0.5  \la z \la 1$, so that
the isothermality may be valid at least up to $z \sim 0.7$. 
Individual extensive modelling of lenses  at $z \sim 1$ also find  
isothermal(-like) total mass distributions of galaxies. 
\citet{SWF07} find a nearly isothermal profile from a non-parametric modelling
of MG0414+054 at $z=0.96$.  \citet{WRK04} model J1632-0033 at $z \sim 1$ and
 find a tight constraint of $\beta=1.91 \pm 0.02$ for a mass density of 
$\rho(r) \propto r^{-\beta}$ ($\beta=2$ corresponding to the isothermal) using
an observed central image. \citet{TK02} combines galactic dynamics and lens 
modelling to find $\beta=2.0 \pm 0.1 ({\rm stat})\pm 0.1 ({\rm syst})$ for
MG 2016+112 at $z=1.004$.   
Moreover, there exist many modelling examples
in which the authors assume the isothermal profile for the lensing galaxies 
at $z \sim 1$ and reproduce the observed image properties although some
systems appear to require substructures. To sum up, strong lensing studies
support the isothermal profile up to $z \sim 1$ without any indication of
 evolution of the total mass profile in redshift. The total mass profiles 
can also be constrained by other dynamical probes although few studies have
been carried out at $z \ga 1$.
\citet{vv08} present velocity dispersion profiles of early-type galaxies 
up to $\sim 4$~kpc ($0''.5$)  at $z \sim 1$ which are consistent with isothermal
profiles. 

Studies based on surface photometry find varying results on the 
evolution of the galaxy mass profiles. 
Many recent studies suggest that high redshift ($z \ga 1$) 
galaxies may be more compact in their light and stellar mass density 
distributions than the local counterparts 
(e.g.\ \citealt{Tru07,vdW08,Cim08,vanD08,Bez09}). In particular,
\citet{Bez09} find that the stellar mass density profiles at $z \sim 2.3$ 
are more steeply declining beyond $r=1$~kpc than the local counterparts. 
\citet{Bez09} also note that the stellar density within 1~kpc
is 2-3 times higher at $z \sim 2.3$ than at $z \sim 0$. This structural 
evolution may be explained by galaxy evolution through  minor mergers 
(\citealt{Naa09,Naa07}). On the other hand, \citet{Hop09} find similar 
densities for
the observed physical radius range both at $z > 2$ and at $z \sim 0$.
\citet{Hop09} further note that there may or may not be differences in the 
densities outside the central regions between $z > 2$ and $z \sim 0$.
The ambiguity outside the central regions arises primarily from the difficulty 
in observing the low surface brightness parts. Interestingly, 
\citet{SLA09} find that early-type galaxies at $1<z<2$ may be divided into 
two groups, one group showing no size evolution and the other group implying 
a significant size evolution compared with local galaxies. These studies
may imply that at least some galaxies undergo size evolutions possibly 
including stellar mass profile evolutions. Complicating the issue, dark 
mass profiles may also evolve as a result of (responding to) stellar mass growth
and dark mass growth via merging. Unfortunately, the baryon-involved evolution 
of dark mass profile is a poorly understood part of galaxy evolution. 
Consequently, the surface photometry data alone cannot
settle the issue of the total mass profile evolution even if the stellar mass 
profile evolution is real. 

As described above direct probes of the galaxy mass profiles give support for
the isothermality up to $z=1$. However, the evidence is not broad and strong
enough to rule out the variation from the isothermality in $z$ and/or $\sigma$ 
(or mass). Hence galaxy mass profiles remain a source of uncertainty. If the 
true mass profiles deviate systematically as a function of $z$ and/or $\sigma$,
then the number density evolution in $z$ as a function of $\sigma$ will be 
biased. Namely, the variation in mass profile leads to the variation in the 
cross section and the magnification bias resulting in a different number density
evolution. Roughly speaking, if the true mass profile is steeper/more 
concentrated (shallower/less concentrated) at $z=1$ than the isothermal profile,
then the number density at $z=1$ should be lower (higher) than that derived
here based on the isothermal profile. It is unclear at present how important 
the effects turn out to be quantitatively. A realistic ray-tracing simulation 
would eventually be required to quantify the effects robustly.

\subsubsection{Local velocity dispersion functions}

The local velocity dispersion function provides a benchmark for the study of
galaxy evolution just as the local luminosity function is. Hence we cannot 
overemphasise the importance of reliable local VDFs for this work. For this
reason we have considered not only the measured early-type VDF by \citet{Cho07}
but also the Monte Carlo simulated VDFs for both the early-type and the 
late-type populations based on the galaxy counts from the SDSS
and the intrinsic correlations between luminosity and internal velocities.

Despite the range of the VDFs considered in this work one may still consider 
the following systematic errors for the local VDFs. First of all, the galaxy 
counts for the type-specific galaxy populations remain somewhat uncertain
 due to the error in the galaxy classifications.
The \citet{Cho07} classification matches relatively well the morphological 
classification. However, even the \citet{Cho07} classification has an error of
about 10\% according to \citet{PC05}.

 Second, the early-type VDF is less reliable 
at lower velocity dispersions ($\sigma \la 150 \kms$) because of
the incompleteness in the volume and magnitude limited samples of galaxies
used by \citet{Cho07} to obtain the early-type VDF and the intrinsic 
correlation between luminosity and velocity dispersion. Indeed, a
Monte Carlo simulated SDSS galaxy sample makes a difference in the
early-type VDF at velocity dispersions $\la 125\kms$ (see Fig.~\ref{VDFs}). 
However, the present strong lensing data are sensitive only to velocity
dispersions $\ga 95\kms$ so that the difference in the VDF for
$\sigma \la 125\kms$ makes little difference in the results on the evolution
of the VDF.

Finally, the late-type VDF is relatively less reliable. We turned the circular
rotation speed to the velocity dispersion assuming the SIS profile 
for all late-type galaxies, but the observed mass profile deviates 
systematically from the isothermal toward the galactic centre. 
In this respect the late-type VDF may not be well-defined and it may be
eventually necessary to use directly the circular velocity function 
for the statistics of strong lensing through a realistic lens model of 
a late-type (spiral) galaxy. Such a realistic modelling will become very 
important in the future when the number statistics of late-type lenses becomes
large. 

\subsection{Comparison with previous results in lensing statistics}

The importance of galaxy evolution in strong lensing statistics was
recognised and applied to a handful of lenses for the first time by 
\citet{Mao91} and \citet{MK94}. However, strong lensing statistics has usually 
been used to constrain the cosmological constant (and dark energy) and 
appears to confirm the concordance cosmology according to recent results 
assuming the early-type population evolves little between $z=1$ and now (e.g.\ 
\citealt{Cha02,Mit05,Cha07,Ogu08}). The assumption of no-evolution
of the early-type population appeared also to be supported by strong lensing
statistics assuming the concordance cosmological model according to an analysis
of the CLASS statistical sample by \citet{CM03} and a lens-redshift test 
by \citet{Ofe03}. Notice that these studies assumed a  constant shape of the
the inferred velocity dispersion function. In other words, \citet{CM03} and
\citet{Ofe03} considered only the evolution of the number density and the
characteristic velocity dispersion. More recently, \citet{CN07} carried out 
the lens-redshift test more extensively and studied in particular possible
systematic errors based on a sample of 70 galaxy lenses. However, \citet{CN07}
also assumed non-evolving shape for the VDF and obtained results consistent
with earlier results. To sum up, all previous results on galaxy evolution
from strong lensing statistics were based on the assumption of the non-evolving
shape for the VDF and appeared to be consistent with no evolution of early-type
galaxies.

In this work we  allow the shape of the VDF to evolve and find that strong 
lensing statistics actually  requires
a differential evolution of the VDF (see Figs.~\ref{Eevol} and
\ref{Aevol}). The differential evolution is intriguing in that the number 
density of galaxies  changes little between $z=1$ and now below 
$\sigma \approx 200 \kms$ but at higher velocity dispersions the number density
 evolution becomes significant with an increasingly larger factor for a higher
velocity dispersion (see Fig.~\ref{Aevol}). 
We discuss the implications of this differential evolution below in \S 6.4.
Here we just point out that this behaviour of differential evolution may 
explain why previous lens statistics resulted in little evolution assuming a 
constant shape. For a constant shape of the VDF its evolution may be realised
only through the vertical and horizontal shifts, namely the variations of the 
integrated number density and  the characteristic velocity dispersion. 
The shifts of the VDF are likely to be determined by the behaviour of the 
number density at the lensing peak velocity dispersion, i.e. the velocity 
dispersion for which the lensing probability ($\propto \phi \sigma^4$) is 
maximal. However, as shown in Fig.~\ref{Aevol}  the number density 
of galaxies evolves little at intermediate velocity dispersions. Consequently,
the VDF appeared to evolve little by previous lensing statistics.

\subsection{Comparison with theoretical predictions and galaxy 
surveys: a qualitative analysis}

In the $\LCDM$ hierarchical structure formation picture the dark halo mass 
function (DMF) evolves in cosmic time as a consequence of hierarchical merging 
(e.g.\ \citealt{WR78,LC93}). Accordingly, the stellar mass function (SMF) and 
the velocity dispersion function (VDF) of galaxies also evolve. However, baryon
physics complicates the evolutions of the SMF and the VDF making it non-trivial
to compare the evolutions of the DMF, the SMF and the VDF one another. 
Conversely, careful analyses of the coevolution of the DMF, the SMF and
the VDF may reveal key insights into galaxy formation and evolution processes
(\citealt{Cha10}). Here we compare the evolutionary trends of the VDF from this 
work against the DMF from N-body simulations and the SMFs predicted by recent
semi-analytical models of galaxy formation and obtained by cosmological surveys
of galaxies. This is meant to be an inter-comparison of the {\it qualitative} 
features of three functions as suggested by current theoretical and 
observational studies. A comprehensive and quantitative analysis of the
coevolution of three functions is given in \citet{Cha10} where evolutions of
functions $\sigma(\Mvir)$ and $\sigma(\Mstars)$ are derived and their
implications are discussed in the context of current cosmological observations.

\begin{figure}
\begin{center}
\setlength{\unitlength}{1cm}
\begin{picture}(8,7)(0,0)
\put(-1.7,8.3){\includegraphics{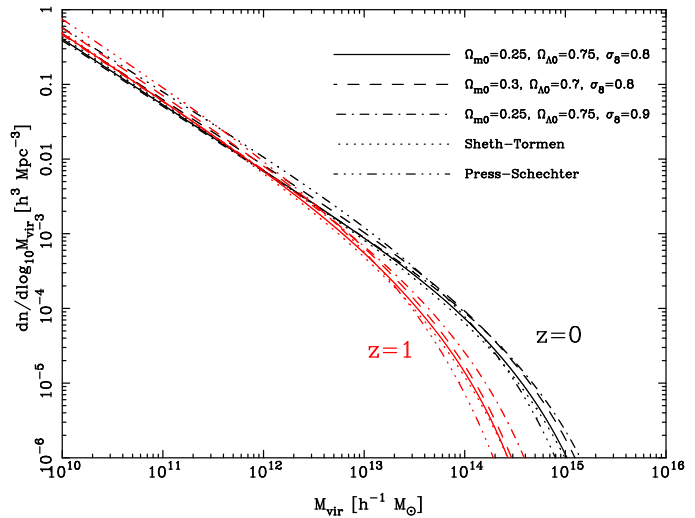}}
\end{picture}
\caption{
The evolution of the dark halo mass function (DMF) between $z=0$ and $z=1$
under the current $\LCDM$ paradigm. The results are based on the code provided
by \citet{Ree07}. Notice that the DMF evolution is differential for 
$\Mvir \ga 10^{13} h^{-1} \Msun$ as is the VDF evolution for 
$\sigma \ga 200 \kms$ (see Fig.~\ref{Aevol}).
}
\label{DMFevol}
\end{center}
\end{figure}

A generic prediction of the $\LCDM$ theory is the differential evolution of the
DMF in which the number density evolution of a larger $\Mvir$ (virial mass) 
halo is greater, e.g.\ since $z=1$ (see Fig.~\ref{DMFevol}). As can be noticed
easily, this evolutionary trend is strikingly similar to that of the total VDF
(Fig.~\ref{Aevol}). In fact, this {\it qualitative} similarity is expected once
we assume that $\sigma$ is on average a monotonically increasing function of
$\Mvir$ at each epoch (a plausible assumption). The quantitative detail of
the evolution of $\sigma(\Mvir)$ has implications on the structural 
evolution of the galaxy-halo system (\citealt{Cha10}).

\begin{figure}
\begin{center}
\setlength{\unitlength}{1cm}
\begin{picture}(9,12)(0,0)
\put(-0.7,0.){\includegraphics{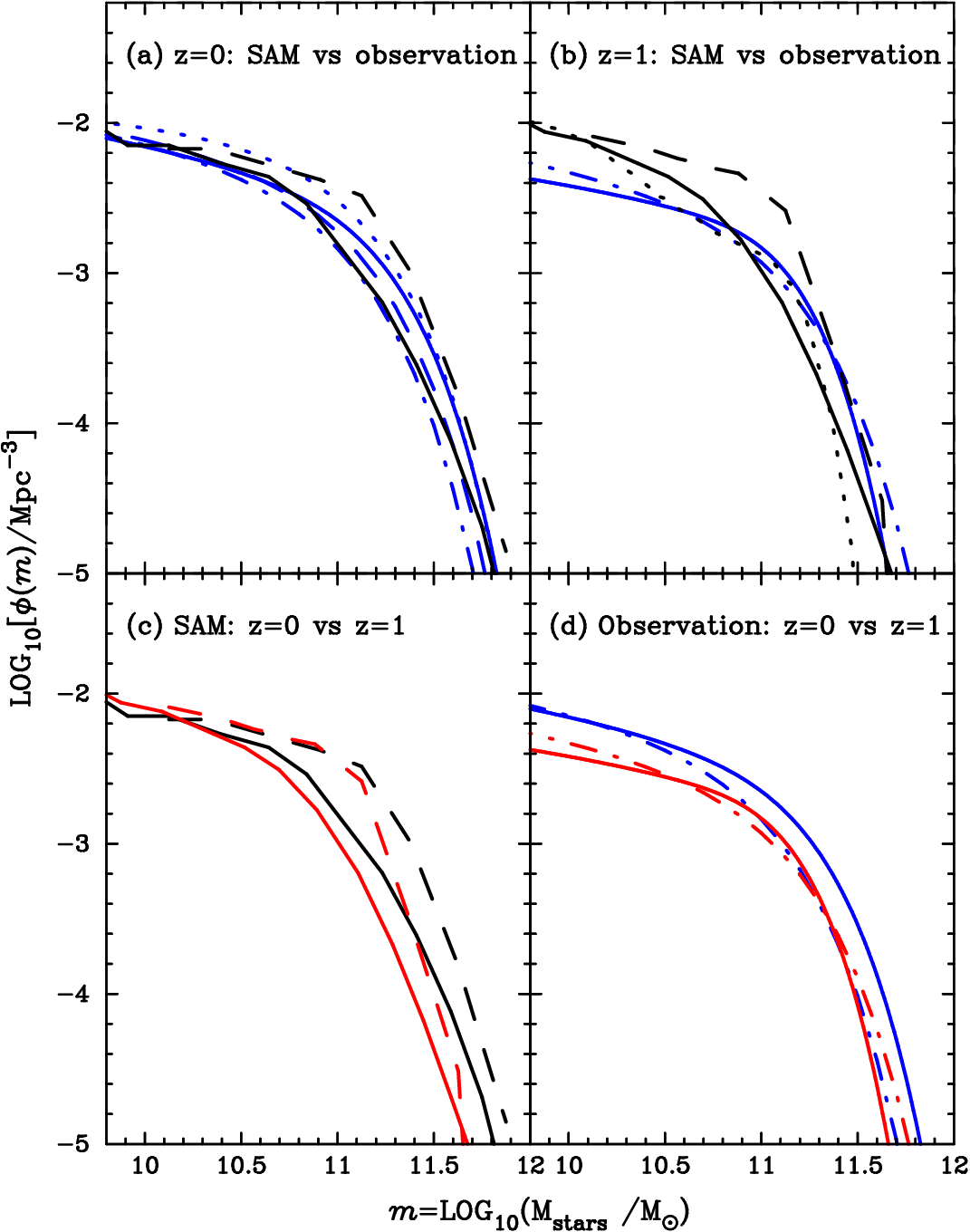}}
\end{picture}
\caption{
(a) Stellar mass functions (SMFs) of galaxies at $z=0$. 
Blue solid, dashed, dash-dotted
 and dotted curves are respectively the observed SMFs from \citet{Ilb09}, 
\citet{Pan07}, \citet{CW09} and \citet{Bel03}. Notice that the  \citet{CW09}
SMF is a composite of several galaxy survey results. Black solid curve is a 
median of the \citet{Fon09} three semi-analytic model (SAM) predictions while 
black dashed curve is the prediction of the \citet{Cat08} SAM. 
(b) SMFs at $z=1$. 
The colours and styles of the lines refer to the same references as in (a).
Black dotted curve is the \citet{Str08} SAM prediction. Notice the discrepancy
between the observed SMFs and the predicted SMFs for the lower part of the
 stellar mass range.
(c) SAM predicted SMFs at $z=0$ and $z=1$ are compared. Black curves refer to 
$z=0$ while red curves refer to $z=1$. Solid and dashed curves are respectively 
the predictions by \citet{Fon09} and \citet{Cat08}. Notice that these 
predicted SMFs show hierarchical differential evolutions in line with the VDF 
and the DMF (see Figs.~\ref{Aevol} and \ref{DMFevol}).
(d) Observed SMFs at $z=0$ and $z=1$ are compared. Blue curves refer to 
$z=0$ while red curves refer to $z=1$.
 Solid and dash-dotted curves are respectively 
the SMFs from \citet{Ilb09} and \citet{CW09}. The composite SMF by \citet{CW09}
shows an anti-hierarchical evolution while the COSMOS SMF by \citet{Ilb09} does
not show such a differential evolution. 
}
\label{SMFevol}
\end{center}
\end{figure}

Unlike the DMF the evolution of the SMF is not well predicted by
theoretical models because of complex galaxy formation processes.
Nevertheless, it appears that recent semi-analytic models (SAMs) predict 
a hierarchical evolution of the SMF that shows evolutionary patterns similar
to those of the DMF and the SMF (see Fig.~\ref{SMFevol}). However, the
`observed' SMFs from galaxy surveys show different evolutionary patterns
(Fig.~\ref{SMFevol}). In particular, many galaxy surveys suggest stellar 
mass-downsizing (apparently anti-hierarchical) evolution of galaxies 
(e.g., \citealt{Cim06,Fon06,Poz07,Con07,Sca07,Coo08,Mar09}),
although there are results that do not particularly support 
stellar mass-downsizing (e.g., \citealt{Bel04,Fab07,Bro08,Ilb09}). 
The disagreement between SAM predictions and galaxy survey results is more
apparent for lower stellar mass galaxies. Some authors (e.g.\ \citealt{Fon09}) 
suggest that this disagreement reflects some additional physical processes 
missing in the current SAMs. However, assuming that $\sigma$ is on average 
a monotonically increasing function of $\Mstars$ at each epoch, 
we may expect a hierarchical evolution of the SMF. 
In this view, the VDF evolution is {\it qualitatively} more in line
with the current SAM predictions  than the 
downsizing SMFs from galaxy surveys. However, if downsizing SMFs from galaxy
surveys should turn out to be true, a deeper rethinking of $\sigma(\Mstars)$
 may be required. A detailed quantitative analysis of the evolution of 
$\sigma(\Mvir)$ based on currently available data is carried out in 
\citet{Cha10}.
 
\section{Summary and conclusions}

In this work we have constrained the evolutionary behaviours of the velocity 
dispersion functions of galaxies based on the local measured and
Monte Carlo realised VDFs and the statistics of 
strongly-lensed radio-selected and optically-selected sources.
This work is also based on the assumption that the total (luminous plus dark)
mass profile is isothermal in the optical region for $0 \le z \le 1$ 
in line with strong lens modelling results.
The constrained evolutions of the VDFs can be characterised as follows.

\begin{itemize}

\item The (comoving) number density of massive galaxies with 
$\sigma \ga 200 \kms$, which are mostly early-type galaxies, evolves 
differentially from $z=1$ to $z=0$ 
in the way that the evolution is greater at
a higher velocity dispersion. The most massive galaxies of 
$\sigma \sim 400 \kms$ are much rarer at $z=1$ than the present epoch
 with its number density probably less than 30\% of the local density.

\item The number density of intermediate-mass galaxies with 
$95 \kms \la \sigma \la 200 \kms$ are nearly constant between $z=1$ and $z=0$
regardless of the morphological type.

\item The velocity dispersion function of late-type galaxies transformed from 
the Monte Carlo realised circular velocity function is consistent with no
evolution between $z=1$ and $z=0$ either in the overall number density or its 
shape. 

\end{itemize}

Comparing the constrained VDFs with the $\LCDM$ DMF and the
SMFs observed from galaxy surveys as well as predicted from 
semi-analytical models of galaxy formation, we find the following.

\begin{itemize}

\item The VDF evolution is in line with the evolution of the $\LCDM$ DMF, i.e.,
 the hierarchical build-up of mass structures over cosmic time.

\item The evolutionary pattern of the VDF is similar to the SMF evolution 
predicted by recent semi-analytical models of galaxy formation but 
qualitatively different from the stellar mass-downsizing evolution suggested
by galaxy surveys. Further investigations are required to clarify this 
apparent conflict.

\end{itemize}

In conclusion we have investigated in detail the evolutionary behaviours of the
VDFs for the first time based on local SDSS data and statistical strong lensing
data and found that the evolution pattern is strikingly similar to that of
the $\LCDM$ DMF. Also, there is a promising agreement 
between the VDF evolution and the SMF evolution predicted by SAMs
although observed stellar mass-downsizing evolution needs yet to be clarified. 
Larger well-defined statistical samples of strong lensing from future
observation tools like the SKA will be invaluable laboratories for decoding
galaxy formation and evolution processes.

\vspace{1cm}
This work makes use of a number of hard-worked observational results from
cosmological surveys including but not limited to the CLASS, the PANELS, the 
HST Snapshot, the SDSS and the 2dF. The author thanks the anonymous referee
for insightful comments and useful suggestions for presentations.
The author received useful comments on the originally submitted version of this
 manuscript from Michael Brown, Lawrence Tresse and Ignacio Trujillo.
The author would like to thank Changbom Park
and Myungshin Im for useful discussions. The author also would like to thank
Shude Mao for comments on the manuscript.

\bibliographystyle{mn2e}

\end{document}